\documentclass[format=acmsmall,natbib=False]{acmart}

\makeatletter
\let\blx@rerun@biber\relax
\makeatother

\RequirePackage[
sortcites,
datamodel=acmdatamodel,
style=acmnumeric, %
natbib=true
]{biblatex} 

\usepackage{listings}
\usepackage{multirow}

\title{Automated Grading and Feedback Tools for Programming Education: A Systematic Review}

\author{Marcus Messer}
\orcid{0000-0001-5915-9153}
\affiliation{\department{Department of Informatics} \institution{King's College London} \city{London} \country{UK}}
\email{marcus.messer@kcl.ac.uk}

\author{Neil C. C. Brown}
\orcid{0000-0001-6086-2479}
\affiliation{\department{Department of Informatics} \institution{King's College London} \city{London} \country{UK}}
\email{neil.c.c.brown@kcl.ac.uk}

\author{Michael K\"olling}
\orcid{0000-0003-0544-2003}
\affiliation{\department{Department of Informatics} \institution{King's College London} \city{London} \country{UK}}
\email{michael.kolling@kcl.ac.uk}

\author{Miaojing Shi}
\orcid{0000-0002-4933-0073}
\affiliation{\department{College of Electronic and Information Engineering} \institution{Tongji University} \city{Shanghai} \country{China}}
\email{mshi@tongji.edu.cn}

\ccsdesc[500]{General and reference~Surveys and overviews}
\ccsdesc[300]{Applied computing~Computer-assisted instruction}
\ccsdesc[500]{Social and professional topics~Student assessment}
\ccsdesc[300]{Social and professional topics~Software engineering education}

\keywords{Automated Grading; Feedback; Assessment; Computer Science Education; Systematic Literature Review; Automatic Assessment Tools}

\usepackage{enumitem}

\newcommand{\skillstyle}[1]{\textbf{\textit{#1}} --}

\begin{document}

\begin{abstract}
   We conducted a systematic literature review on automated grading and feedback tools for programming education.
   We analysed 121 research papers from 2017 to 2021 inclusive and categorised them based on skills assessed, approach, language paradigm, degree of automation and evaluation techniques.
   Most papers assess the correctness of assignments in object-oriented languages.
   Typically, these tools use a dynamic technique, primarily unit testing, to provide grades and feedback to the students or static analysis techniques to compare a submission with a reference solution or with a set of correct student submissions.
   However, these techniques' feedback is often limited to whether the unit tests have passed or failed, the expected and actual output, or how they differ from the reference solution.
   Furthermore, few tools assess the maintainability, readability or documentation of the source code, with most using static analysis techniques, such as code quality metrics, in conjunction with grading correctness.
   Additionally, we found that most tools offered fully automated assessment to allow for near-instantaneous feedback and multiple resubmissions, which can increase student satisfaction and provide them with more opportunities to succeed.
   In terms of techniques used to evaluate the tools' performance, most papers primarily use student surveys or compare the automatic assessment tools to grades or feedback provided by human graders.
   However, because the evaluation dataset is frequently unavailable, it is more difficult to reproduce results and compare tools to a collection of common assignments.
\end{abstract}

\maketitle

\section{Introduction}\label{sect:intro}
Most computer science courses have grown significantly over the years, leading to more assignments to grade \cite{Krusche2020}.
The time window for evaluating assignments is typically short as prompt and timely feedback increases student satisfaction \cite{Kane2008}, resulting in the danger of inconsistent grading and low feedback quality.

One method for providing a grade and feedback in good time is to use multiple human graders.
This approach, however, increases the chance of variation in grading accuracy, consistency and feedback quality \cite{Aziz2015}.  As a result, automatic grading tools have become widely popular, as they are able to assign grades and generate feedback consistently for large cohorts.
Automatic Assessment Tools (AATs) may be used by instructors either to fully automate the marking and feedback process, or to indicate potential issues while manually assessing the submissions.

There is typically a relationship between grading and feedback when assessing student submissions for a given assignment.
Formative assessment focuses on providing feedback to teachers and students to help students learn more effectively while providing an ongoing source of information about student misunderstanding \cite{Dixson2016}.
Whereas summative assessments typically intend to capture what a student has learned and judge performance against some standards and are almost always graded \cite{Dixson2016}.
While some AATs focus on providing only feedback for formative assessments, typically, many provide a grade and feedback for summative assessments.
In this paper, we will refer to providing a grade and/or feedback as an assessment unless we explicitly discuss grading or feedback exclusively.

Using AATs to grade programming assignments originated in the 1960s \cite{Hollingsworth1960}. 
Traditionally, AATs have focused on grading program correctness using unit tests and pattern matching. 
They typically struggle, however, to grade design quality aspects and to identify student misunderstandings.
Only recently have researchers begun to address these areas as well \cite{Orr2021, Ureel2019}.

As part of implementing a unit test-based AAT, instructors must provide a comprehensive test suite. 
It typically takes considerable effort and requires students to follow a specific structure when implementing their solutions, such as replicating the exact output being tested or using predefined class and function names.
The considerable effort and specific structure often lead to instructors designing short and well-defined coursework, as it is easier to implement a comprehensive test suite for such assignments.
However, some instructors prefer to incorporate large-scale project-based assignments in their courses, which are typically less well-defined to allow students to control the direction of the assignment and use their creativity, both of which can motivate students to perform well in their assignments.
These open-ended assignments are usually nearly impossible to automatically assess using unit test-based approaches, as the structure and functionality of the student assignments can change, making it difficult to implement unit tests.

We conducted a systematic literature review to investigate recent research into automated grading and feedback tools.
Our review offers new insights into the current state of auto-grading tools and the associated research by contributing the following:

\begin{itemize}
    \item Categorised automatic assessment tools based on core programming skills \cite{Messer2023}.
    \item A summary of the state of the art. %
    \item Detailed statistics of grading and feedback techniques, language paradigms graded, and evaluation techniques.
    \item An in-depth discussion of the gaps and limitations of current research.
\end{itemize}

We utilised the programming skills and machine learning (ML) papers from the results of this systematic literature review to further investigate ML-based AATs by performing a meta-analysis focusing on ML-based AATs.
In the meta-analysis \cite{Messer2023}, the ML papers from the current review were used as initial papers for a backward snowball search to find other ML-based AATs.
After conducting the snowball search and finalising our included papers, we categorised the ML-based AATs using the core skills discussed in Section \ref{sect:skills}, and the techniques and evaluation criteria utilised.

The format of the article is as follows: Sections \ref{sect:skills} and \ref{sect:cats} introduce our framework for categorising AATs.
Section \ref{sect:related} presents existing reviews and discusses how our own work relates to and expands on this prior work.  Section \ref{sect:method} details our methodology, inclusion and exclusion criteria and introduces the research questions addressed by this work.
Section \ref{sect:results} presents an analysis of the selected sources and presents the results, and Section \ref{sect:discussion} discusses our findings.
Finally, Section \ref{sect:conculusion} summarises our conclusions and presents recommendations.

\section{Programming Skills}\label{sect:skills}
We distinguish four core criteria for assessing programming assignments based on our experience: correctness, maintainability, readability, and documentation.
These criteria are manually or automatically graded by evaluating a student's source code statically or dynamically and have been active research areas within computer science education.

There are multiple facets of correctness research, including students' conceptions and perspectives of correctness \cite{Kolikant2008, Ioanna2006, Qian2018}, how students fix their code to improve correctness \cite{Souza2017, Alqadi2017}, and assessment of correctness \cite{rayyan-354359303, rayyan-354359334, Gerdes2010}.
Furthermore, there has been a multitude of research into code quality, which includes readability and maintainability \cite{Borstler2018, Stamelos2002, Kanellopoulos2010}.
The research has included designing rubrics for grading code quality \cite{Stegeman2016} and investigating code quality issues within student programs \cite{Keuning2017}.
Finally, code documentation is considered one of the best practices in software development \cite{Kipyegen2013, Souza2005, Rai2022} and is widely researched within the software engineering domain, including code summarisation \cite{McBurney2015, Haque2022}, how documentation is essential to software maintenance  \cite{Souza2005} and how to prioritize documentation effort \cite{McBurney2018}.
We separated code assessment into these four skills to provide a comprehensive overview of which skills are automatically graded.

The four core criteria provide our frame of reference for investigating research into AATs for evaluating student submissions:

\skillstyle{Correctness}
Evaluate whether a student has understood and implemented the tasks in a manner that conforms to the coursework specification.
There are two commonly evaluated areas of correctness.
The most common is correctness of functionality: testing whether a student has adequately implemented the assignment's essential features, whether correctly calculating the Fibonacci sequence, implementing \texttt{FizzBuzz}, or creating a text-based adventure game.
The other is the correctness of methodology, verifying that a student has used a specific language feature, such as recursion to calculate the Fibonacci sequence, a for-loop and modulo to implement \texttt{FizzBuzz}, or polymorphism and inheritance to create an object-oriented game.

\skillstyle{Maintainability}
Investigates how well a student has implemented maintainable or elegant code.
This could include how well a student has used functions to reduce duplicated code, whether the method used to solve a problem is simplistic or overly complicated or if the student has used polymorphism and inheritance to reduce class coupling.

\skillstyle{Readability}
Analyses whether a student's submission is easy to understand.
While maintainable code contributes to the readability of the code, other aspects of source code can signify if the code is readable.
These aspects include: following code style guidelines, meaningful naming of classes, functions and variables, replacing magic numbers with constants, and using whitespace to separate code blocks.

\skillstyle{Documentation}
Inspects the existence and quality of a student's documentation, including inline comments and docstrings.
While some believe code should be self-documenting, including inline comments to explain functionality and implementation is common practice.
However, these are often useless \cite{Raskin2005} or poor quality \cite{Steidl2013}.
Additionally, developers must include documentation in the form of docstrings to explain the purpose and the interactions for a specific class or function. 
These are typically used when other developers interact with or maintain the function \cite{Steidl2013}.

\newpage
\section{Categories of Automatic Assessment Tools}\label{sect:cats}
AATs have numerous advantages for both educators and students.
They typically reduce the time it takes to assess each assignment and can be used to enhance the traditional assessment workflow and establish a real-time feedback system.
The process of automatically assessing source code can be done in various ways, such as unit testing and comparing source code to an instructor's model solution or other students' correct submissions.
Depending on the assignment's intended learning outcomes, an AAT may employ multiple methods to grade various core skills.

In 2005, \citeauthor{Mutka2005} published a ``A Survey of Automated Approaches for Programming Assignments'', where they categorised automatic assessment of different features into two main categories: dynamic analysis and static analysis \cite{Mutka2005}.

\subsection{Dynamic Analysis Automatic Assessment Tools}
Dynamic analysis evaluates a running program's attributes by determining which properties will hold for one or more executions  \cite{ball1999}.
Typical methods employed by dynamic analysis AATs include using a suite of unit tests to grade the correctness of a student's submission by comparing either the printed output or return values of individual methods.

Additionally, dynamic analysis can be used to evaluate the student's ability to write efficient code or to write complete test suites \cite{Mutka2005}.
To create a comprehensive test suite, instructors typically need to be skilled at creating unit tests using complex language features such as reflection and invest significant time developing and evaluating the tests. 
Test suites are usually given to students in one of three ways: complete access before the final deadline \citep{rayyan-354359326,rayyan-354359338,rayyan-354359344}, no access before the final deadline \citep{rayyan-354359314,rayyan-354359318,rayyan-354359305,rayyan-354359300,rayyan-354359310}, or access to a subset of the test suite before the final deadline \citep{rayyan-354359276,rayyan-354359307,rayyan-354359353}.
How the test suites are presented to students depends on the assessment approach.
Typically, if the assignment is formative, students will be given complete access throughout; if the assignment is summative, the students will either have no access or only access to a subset of tests.
Limiting students' access to the complete set of tests or using hidden tests are typically used to limit the students' ability to game the system by hardcoding return values.

\subsection{Static Analysis Automatic Assessment Tools}
Static analysis tools assess software without running it or taking inputs into account \cite{Ayewah2008}.

\subsubsection{Static Analysis Tools}
Many industry-standard static analysis tools have been implemented into AATs, including linters, such as pylint\footnote{pylint -- a Python Linter:  \url{https://pypi.org/project/pylint/}} and cpplint\footnote{cpplint -- A C++ Linter: \url{https://github.com/cpplint/cpplint}}, CheckStyle\footnote{CheckStyle --  A style guide enforcement utility: \url{https://checkstyle.sourceforge.io/}}, FindBugs\footnote{FindBugs has since been abandoned, and replaced with SpotBugs: \url{https://spotbugs.github.io/}} \cite{Ayewah2008}, and PMD\footnote{PMD -- A static analyser focusing on finding programming flaws: \url{https://pmd.github.io/}}. 
These static analysis tools typically focus on identifying or partially identifying issues, and checking maintainability,  readability, and the existence of documentation.

\subsubsection{Software Metrics}
Software metrics are a technique for assessing code most commonly used in commercial settings that have been effectively integrated into AATs \cite{Mutka2005}.
Metrics are typically used to evaluate maintainability; \citeauthor{Halstead1977} and \citeauthor{Mccabe1976} created such complexity measures.
\citet{Halstead1977}'s metrics include program length, comprehension difficulty, and programming effort. whereas \citet{Mccabe1976}'s Cyclomatic Complexity utilises graph theory to evaluate the program's complexity based on the control flow graph.
Other than complexity, metrics can be used to evaluate object-oriented design principles, such as class coupling and depth of inheritance \cite{Chidamber1994}.

\subsubsection{Comparative Automatic Assessment Tools}
Other than adapting industry tools, static analysis has been used to compare a student's submission with a model solution or to a set of applicable solutions \cite{Mutka2005}, typically focusing on grading correctness.
These tools use various methods to evaluate the similarity between the student's submission and the model solution(s).
Source code can be written differently but implement the same functionality.
To improve accuracy, AATs can convert the source code to an abstract syntax tree \cite{Wang2020} or a control flow graph \cite{Sendjaja2021} to abstract away from the syntactic representation.
More comprehensive AATs have multiple methods for matching or partially matching student solutions to the model solution(s).
Virtual Teaching Assistant \cite{Chou2021} incorporates four patterns to facilitate grading partial correctness: location-free, where the order of output tokens or characters or specific structure is not a determining factor in the awarded grade and location-specific, where the order of the output or the specific structure does impact the awarded grade.

Comparative approaches can also be used to generate feedback for the student.
\citet{Paassen2017} use edit-based approaches on student trace data to generate next-step hints for block-based programming languages. %

\subsection{Machine Learning Automatic Assessment Tools}
Since \citet{Mutka2005} published their review in 2005, research into applying machine learning to auto-grading has increased.
Machine learning AATs use various techniques to grade or provide feedback on the correctness and maintainability of a student's submission, typically utilising dynamic or static approaches.
\citet{Orr2021} trained a feed-forward neural network to grade the design quality and provide personalised feedback.
They converted the source code to an abstract syntax tree and then converted it to a feature vector as the input to the regression model that predicts the score.

To grade correctness, \citet{Dong2020} implemented two ML AATs into an online judge.
The first model was trained using historical training data to predict what causes failed test results.
The second model implemented was a knowledge-tracing model used to predict the probability of a student passing a new problem based on previous knowledge of a programming concept.

However, the approaches of both \citeauthor{Orr2021}, and \citeauthor{Dong2020}  require a large ground truth dataset to train their models.
A zero-shot learning approach can resolve the need for large training data. 
\citet{Efremov2020} implemented such an approach to provide next-step hints to students where no prior historical data for a task exists.
They implemented a Long Short-Term Memory (LSTM) neural network to provide a vector representation that can be used for ML of a block-based language, followed by a reinforcement learning approach for the hint policy.

\subsection{Degree of Automation}
\citet{Mutka2005} also defined varying degrees of automated the assessment of programming assignments.
Fully automated assessment is typically used for smaller assignments where unit testing or another form of automatic grading can easily be designed and implemented, such as assignments focusing on programming language basics \cite{Mutka2005}.
Large courses use fully automated grading to reduce the workload on the instructors.
However, fully automated grading tools have difficulty grading large-scale assignments, graphical user interfaces, maintainability, readability, and documentation, which are typically manually graded.

It is common to use a semi-automated approach, a mix of manual and automated assessment, to minimise the instructors' workload on these large-scale assignments.
Automating certain aspects of the assessment process allows the instructor more time to grade and give feedback on areas that cannot be easily automated, including code design \cite{Mutka2005}.

\section{Related Work}\label{sect:related}
Multiple studies have reviewed the literature and existing automatic assessment tools for programming assignments. 
These survey papers focused on literature that discussed grading assignments or feedback given to students on their work.

\subsection{Grading}
Most recently, \citet{Paiva2022} reviewed which computer science domains were automatically assessed.
They reviewed proposed testing techniques, how secure code execution is, feedback generation techniques, and how practical the techniques and tools are. 

\citeauthor{Paiva2022}'s review showed that most automated assessment research focused on multiple programming domains, including visual programming, web development, parallelism and concurrency.
Additionally, they found that the tools and techniques for assessing assignments are in one or more categories: functionality, code quality, software metrics, test development, or plagiarism.
When evaluating feedback, they used \citet{Keuning2018}'s feedback categories (discussed in Section \ref{sect:lit:feedback}). A novel feedback approach they found was using automated program repair to fix software bugs.

Additionally, \citeauthor{Paiva2022} classified each tool into web-based platforms, Moodle plug-ins, web services, cloud-based services, toolkits and Java libraries and analysed the application of the found techniques to the tools.
Finally, they discussed the previous and future trends of the key topics; the most notable previous key topics include tool development, static analysis and feedback.

We previously introduced \citet{Mutka2005}'s survey on automated approaches for programming assignments, where they surveyed the literature to define different types of automated assessment approaches.
They discussed the benefits and drawbacks of automated assessment and encouraged careful assignment design to provide students with several practical programming tasks.

Furthermore, \citet{Ullah2018} expanded upon \citeauthor{Mutka2005}'s work by introducing a hybrid category consisting of AATs that are both dynamic and static approaches and investigating automated assessment tools released before the paper's publication in 2018. 
They introduce a taxonomy of approaches for both static and dynamic approaches and discuss the approach, supported languages, advantages and limitations of existing tools.

\citet{Aldriye2019} analysed a small set of existing grading systems in multiple areas, including usability, understandability of system feedback, and the advantages compared to other tools.
Similarly, \citet{Nayak2022} discussed the implementation and functionality of numerous automated assessment tools, and \citet{Lajis2018} reviewed AATs which used semantic similarity to a model solution to grade submissions.

\citet{Douce2005}, conducted a literature review of the most influential AATs from the earliest example in 1960 by \citet{Hollingsworth1960}, until the paper's publication in 2005, and discuss the change of AATs throughout time, from early assessment systems, to tool-oriented systems, finally to web-based tools.
Whereas, \citet{Ihantola2010} investigated the approaches tools from 2006 to 2010 utilised both from a pedagogical and technical point of view, including programming languages assessed, how instructors define tests and if any tools specialised in specific areas such as assessment of GUIs.

\citet{Souza2016} conducted a systematic literature review of AATs between 1985 and 2013, focusing on key characteristics of AATs, including supported programming language, user interfaces and types of verification.
They categorise the tools by degree of automation, if the tool is instructor or student-centred, and if the tools are specialised, such as contents, testing or quizzes.

\subsection{Feedback}\label{sect:lit:feedback}
\citet{Keuning2018} conducted a systematic literature review of automated programming feedback.
They aimed to review the nature of feedback generated, the techniques used, the tools' adaptability and the quality and effectiveness of the feedback.

To categorise the feedback types, \citeauthor{Keuning2018} used and built on \citeauthor{Narciss2008}'s \cite{Narciss2008} feedback components; these include: knowledge of performance for a set of tasks, knowledge of result/response, knowledge of correct results, knowledge about task constraints, and four others.
They found that the most common feedback types were knowledge about mistakes and how to proceed.

To evaluate how teachers can adapt the tools, \citeauthor{Keuning2018} categorised the tools by the different types of input teachers can provide, including model solutions, test data and solution templates.
Finally, they investigated how authors evaluated the quality and effectiveness of the feedback or tool and found that some completed most evaluations for technical analysis, such as comparing generated grades or feedback with an existing dataset of graded work.
Other evaluation techniques include anecdotal evidence, surveys and learning outcome evaluations.

\subsection{Novelty of This Review}
Multiple reviews are small non-systematic reviews that hand-picked tools to assess and summarise \cite{Douce2005, Lajis2018, Aldriye2019, Nayak2022, Ullah2018}.
Our review differs from these by systematically searching the literature and extracting more detail about each system.

While \citet{Keuning2018} investigated feedback generation, which overlaps with half of our review, their review only considered papers up to 2015 inclusive.
Thus, none of their included papers overlap with ours.
Nevertheless, their results form a valid distinct comparison with ours. 
Similarly, \citet{Ihantola2010} and \citet{Souza2016}, investigated automated grading systems between 2006 and 2010 and is now outdated, given the recent growth in automated assessment systems.

\citet{Paiva2022} carried out a state-of-the-art review in 2022 concurrently with this work.
Several of their research questions (such as code execution security) have no overlap with this review.
They did investigate which aspects of programs are assessed, including quality and the techniques used to generate feedback.
However, they did not cover the evaluation of these tools in detail, in contrast to our detailed examination of evaluation versus human graders.
Their review focuses more on the technical aspects of the automated graders. 
In contrast, ours takes a more pedagogical approach to consider the automated graders' place within education, focusing on how well they grade, what they grade and what they should be grading.

To summarise, our review introduces a new method of categorising AATs by the core skills graded, whether that be correctness, maintainability, readability or documentation, and builds on \citet{Mutka2005}'s categories for approaches to assessing programming assignments, by introducing a category for machine learning AATs, which is a sub-category of both static and dynamic approaches.
We also include an analysis of the evaluation techniques used and the data availability and an analysis of the techniques used to grade or give feedback on the source code.
In Section \ref{sect:discussion:related_work}, we further compare the results of our literature review to previous literature reviews.

\section{Methodology}\label{sect:method}
A systematic literature review is a method for locating, assessing, and interpreting all accessible data on a specific subject.
They are frequently used to summarise existing evidence, identify gaps, and set the stage for new research endeavours \citep{Kitchenham2007}.
Conducting a systematic review involves developing a review protocol, identifying research, selecting primary studies, study quality assessment, data extraction and data synthesis.

This section discusses our review protocol, defining our research questions, inclusion and exclusion criteria, and search and screening processes.
We used Rayyan \cite{Ouzzani2016} to aid our screening process, using their in-built tools for automated and manual de-duplication and keyword highlighting derived from our search criteria.
We did not use their paper clustering feature to aid the screening process.
Section \ref{sect:results} discusses the outcomes of selecting primary studies, quality assessment, data extraction and data synthesis.

\subsection{Preregistration}
We preregistered our study \cite{Messer2022} with the Open Science Foundation.
In the preregistration, we provided a complete description of our planned methodology, including our research questions, search terms, screening, data extraction, and data synthesis.

\subsection{Research Questions}
To guide our review, we used the following research questions:

\begin{enumerate}[label={\textbf{RQ\arabic*}}, align=left]
    \item Which are the most common techniques for programming automated assessment tools?
    \item Which programming languages do programming automated assessment tools target?
    \item Which critical programming skills, such as correctness, maintainability, readability and documentation, are typically assessed by programming automatic assessment tools?
    \item How well do the automated grading techniques perform compared to a human grader?
    \item Is the feedback generated by these techniques comparable to human graders?
    \item What are the most common methods for providing feedback on a student's programming assignment, and what areas do they address?
 \end{enumerate}

\subsection{Search Strings}
We defined two search strings, one for automated grading (Listing \ref{lst:grading}) and the other for automated feedback (Listing \ref{lst:feedback}).
These were defined by extracting keywords from our research questions.
After finding the keywords, we extended our search string to include the keywords' synonyms.
To find the variants of a keyword, we stemmed the keywords and added a wildcard character.
For example, ``grading'' becomes ``grad*''.
However, the stemming method yielded numerous irrelevant results (e.g. gradient) due to the common nature of our keywords.
Therefore, we decided to use specific variants, such as ``grade'', ``grading'' and ``grader''.

We decided to include a negation statement in our search strings to reduce the number of outside sources, which eliminates sources that focus on robotics, source code vulnerability, or information and communication technology, as these are out of the scope of this review. 

Our search strings have minor differences to maximise the likelihood of our benchmark papers being located in our database search.
In Listing \ref{lst:feedback}, `student code' and `novice programm*' were included to allow the search to return the benchmark papers \cite{Piech2015, Parihar2017, Singh2013}.
While the benchmark papers chosen for Listing \ref{lst:grading} did not include the terms `student code' or `novice programming' in their title or abstracts.

\begin{minipage}{\linewidth}
    \begin{lstlisting}[columns=fullflexible,caption=Grading Search String, captionpos=b, label=lst:grading]
        (programming OR source code) AND 
        (grade OR grading OR grader OR 
            mark OR marks OR marking) AND 
        (assignment OR exercise OR assessment OR course) AND
        NOT(robot* OR vulnerability OR ICT)
    \end{lstlisting}
    
    \begin{lstlisting}[columns=fullflexible,caption=Feedback Search String, captionpos=b, label=lst:feedback]
        (programming OR source code OR student code) AND
        (feedback OR hint) AND
        (assignment OR exercise OR 
            submission OR novice programm*) AND
        NOT(robot* OR vulnerability OR ICT)
    \end{lstlisting}
\end{minipage}

\subsection{Search Process}\label{sect:method:search}
To locate a set of relevant primary studies, we used our search strings in the ACM Digital Library (ACM DL), IEEEXplore, and Scopus.
We chose these databases since ACM or IEEE publishes most Computer Science Education resources.
We used Scopus, the world's largest abstract and citation database for peer-reviewed literature, to find additional sources outside of ACM and IEEE.
 
Initially, we investigated other databases, specifically Google Scholar, ScienceDirect, and SpringerLink.
Google Scholar and SpringerLink, returned almost 400,000 results, and ScienceDirect did not support the number of boolean terms in our search strings.
We could not minimise the number of results by restricting our search strings because ``feedback'', ``mark'', ``assessment'', and ``code'' are terms in many other domains, including medical and genetics publications, and these databases do not allow for filtering by domain.   Thus we did not use these databases.

We implemented a benchmark using papers we found during our initial reading to validate our search strings by confirming that the search results contain known papers that we identified during our initial reading.
For automated grading we used five papers \citep{Douce2005, Hollingsworth1960, Insa2015, Parihar2017, Rahman2020}, and for automated feedback we used seven papers \citep{Parihar2017, Singh2013, Gusukuma2018, Denny2021, Keuning2018, Watson2012, Piech2015}.
The results from our benchmark can be found in Table \ref{tab:benchmark} and was performed without applying our inclusion/exclusion criteria (Table \ref{tab:criteira}).
As shown in Table \ref{tab:benchmark}, IEEEXplore does not contain any of our benchmark sources, as IEEE did not publish any of our benchmark sources.
Even though there are no benchmark sources, our search strings should still produce relevant results in IEEEXplore, as the search strings provide relevant sources from ACM and Scopus.

\begin{table}[b]
    \footnotesize
    \begin{tabular}{|r|cc|cc|}
    \hline
    \multirow{2}{*}{\textbf{Database}} & \multicolumn{2}{c|}{\textbf{Grading Search String}}       & \multicolumn{2}{c|}{\textbf{Feedback Search String}}      \\ \cline{2-5} 
                                       & \multicolumn{1}{c|}{Found} & \multicolumn{1}{|c|}{Total}  &\multicolumn{1}{c|}{Found} & \multicolumn{1}{c|}{Total} \\ \hline
    ACM DL                             & \cite{Insa2015, Douce2005, Hollingsworth1960, Parihar2017} & \multicolumn{1}{|c|}{4}& \cite{Singh2013,Denny2021,Keuning2018} & \multicolumn{1}{|c|}{3} \\ \hline
    IEEEXplore                         & - & \multicolumn{1}{|c|}{0} & - & \multicolumn{1}{|c|}{0}\\ \hline
    Scopus                             & \cite{Douce2005, Hollingsworth1960} & \multicolumn{1}{|c|}{2} & \cite{Singh2013, Gusukuma2018, Denny2021, Keuning2018, Watson2012, Piech2015} & \multicolumn{1}{|c|}{6} \\ \hline
    \textbf{Total}                     & \multicolumn{2}{c|}{\textbf{4/5}}                         & \multicolumn{2}{c|}{\textbf{6/7}}                         \\ \hline
    \end{tabular}
    \caption{Benchmark Results}
    \label{tab:benchmark}
\end{table}

\subsection{Inclusion/Exclusion Criteria}\label{sect:method:criteria}
We employed the inclusion and exclusion criteria listed in Table \ref{tab:criteira} to determine whether a primary study is eligible for our review.
We decided only to include sources supporting textual languages, such as Java,  Python and assembly. 
We only included sources from 2017--2021 inclusive to ensure we focused on the most up-to-date research.
We omitted sources we could not access or those not written in English because we could not extract the required information.
Similarly, we rejected posters and brief articles since they are unlikely to have sufficient detail.
Finally, we removed non-peer-reviewed studies because the quality and veracity of these sources could not be verified. 

\begin{table}[t]
\footnotesize
    \centering
    \begin{tabular}{|p{0.4\textwidth}||p{0.4\textwidth}|}
        \hline
        \multicolumn{1}{|c|}{\textbf{Inclusion}} & \multicolumn{1}{|c|}{\textbf{Exclusion}}\\ \hline
        \multicolumn{1}{|p{0.4\textwidth}|}{The paper is a primary source.} & \multicolumn{1}{|p{0.4\textwidth}|}{The paper is not written in English.} \\ \hline
        \multicolumn{1}{|p{0.4\textwidth}|}{The paper focuses on auto-grading or feedback on source code/programming assignments.} & \multicolumn{1}{|p{0.4\textwidth}|}{The paper is not a peer-reviewed paper.}\\ \hline
        \multicolumn{1}{|p{0.4\textwidth}|}{The tool supports a textual programming language.} & \multicolumn{1}{|p{0.4\textwidth}|}{The paper is not accessible via university subscriptions.}\\ \hline
        \multicolumn{1}{|p{0.4\textwidth}|}{The paper was published in 2017 to 2021 inclusive.} & \multicolumn{1}{|p{0.4\textwidth}|}{Posters and papers shorter than four pages.}\\ \hline
        \multicolumn{1}{|p{0.4\textwidth}|}{} & \multicolumn{1}{|p{0.4\textwidth}|}{Tools that use visual/block-based programming languages.}\\ \hline
    \end{tabular}
    \caption{Inclusion \& Exclusion Criteria}
    \label{tab:criteira}
\end{table}

\subsection{Screening Process}\label{sect:method:screening}
We used two screeners to conduct our screening process, with any conflicts between decisions being discussed and resolved at regular intervals.
We included or excluded studies based on their title and abstract for our first screening stage.
In the second stage, we screened the introduction and conclusion.
Finally, we screened the full-text sources and extracted data. 

We changed our screening approach slightly (for the better) from our proposed approach in our preregistration \cite{Messer2022}.
We decided that both screeners would review all sources during all stages of the review, as this would increase the reliability of the results.

\section{Results}\label{sect:results}
\begin{figure}[]
      \centering
    \includegraphics[width=0.25\linewidth]{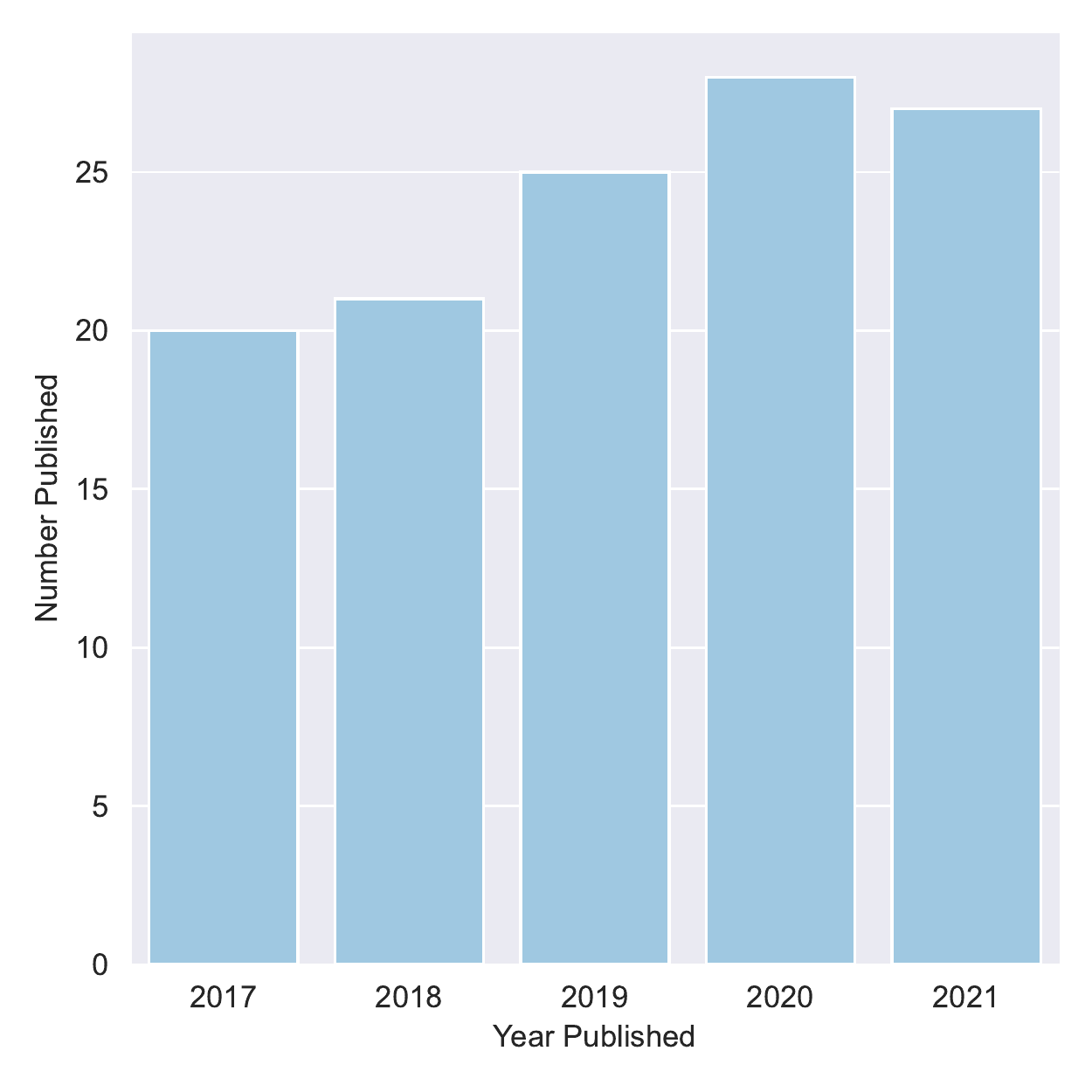}
    \caption{The number of papers published per year.}
    \label{fig:years_published}
\end{figure}

Figure \ref{fig:years_published}, shows the general trend of automatic assessment-related papers over time.
The increasing trend of research into automated grading correlates to the increasing class sizes. 

\begin{figure}
    \centering
    \includegraphics[width=0.55\linewidth]{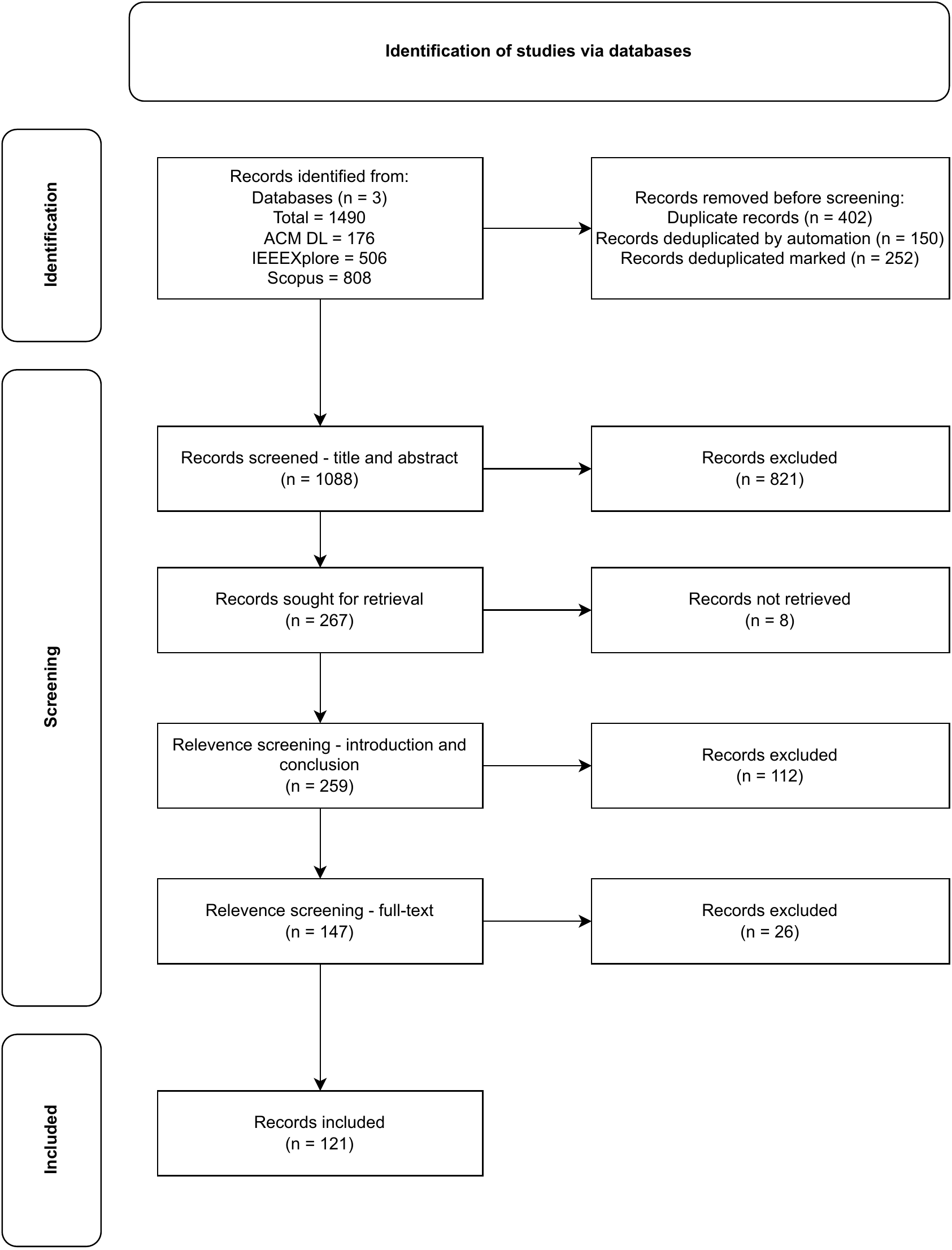} 
    \caption{PRISMA \cite{Page2021} diagram showing the inclusion and exclusion decisions at each stage.}
    \label{fig:prisma}
\end{figure}

Figure \ref{fig:prisma} shows an overview of the number of included and excluded papers in each stage of the review process.
Our search for primary sources resulted in 1490 papers from the three databases, which after automated and manual deduplication, resulted in 1088 papers to screen by title and abstract.
We excluded 820 publications based on our exclusion criteria during the title and abstract screening stage. 
Most of the exclusions were due to the paper's lack of focus on grading or feedback on programming assignments.

To conduct our introduction and conclusion screening, we sought the full text of 268 papers, 8 of which we could not retrieve due to a lack of access to the publication.
Of the 260 papers we retrieved, 112 were excluded, and most were excluded for not focusing on grading or feedback, paper length, or focusing on visual programming languages.

In the full-text stage of our screening, we reviewed 147 articles.
Out of the 147, 26 were excluded, most of which we excluded as they worked towards a grading or feedback tool rather than discussing a completed tool, leaving 121 papers to extract data from.
The final included papers can be found in the supplementary material grouped by research question.

We opted to annotate and analyse the papers themselves instead of extracting each tool from the articles, as the aim of this literature review is to provide an overview of current state-of-the-art research.
Very few papers discussed multiple tools, and for these papers, we annotated based on all the tools discussed, which were typically within the same domain as each other.
Furthermore, few tools were discussed in multiple papers, with only  Antlr \cite{rayyan-354359280, rayyan-354359323, rayyan-354359325, rayyan-354359332, rayyan-354359360}, VPL \cite{rayyan-354359269, rayyan-354359370, rayyan-354359376}, Coderunner \cite{rayyan-354359282, rayyan-354359337, rayyan-354359349} and Travis \cite{rayyan-354359293, rayyan-354359363, rayyan-354359385} appearing in three or more papers, with Antlr and Travis primarily being underlying technologies for AATs.
While there is some duplication in terms of papers discussing multiple tools and tools being discussed in multiple papers, this duplication does not overly skew the results of the literature review.

\subsection{Skills Evaluated and Utilised Techniques (RQ1, RQ3, RQ6)}
In this section, we present our results for the techniques utilised to automatically assess student assignments.
We categorise the tools on the previously defined core programming skills as correctness, maintainability, readability and documentation (Section \ref{sect:skills}), and the categories of AATs, including the degree of automation (Section \ref{sect:cats}).

\begin{figure}
        \centering
        \includegraphics[width=0.6\textwidth]{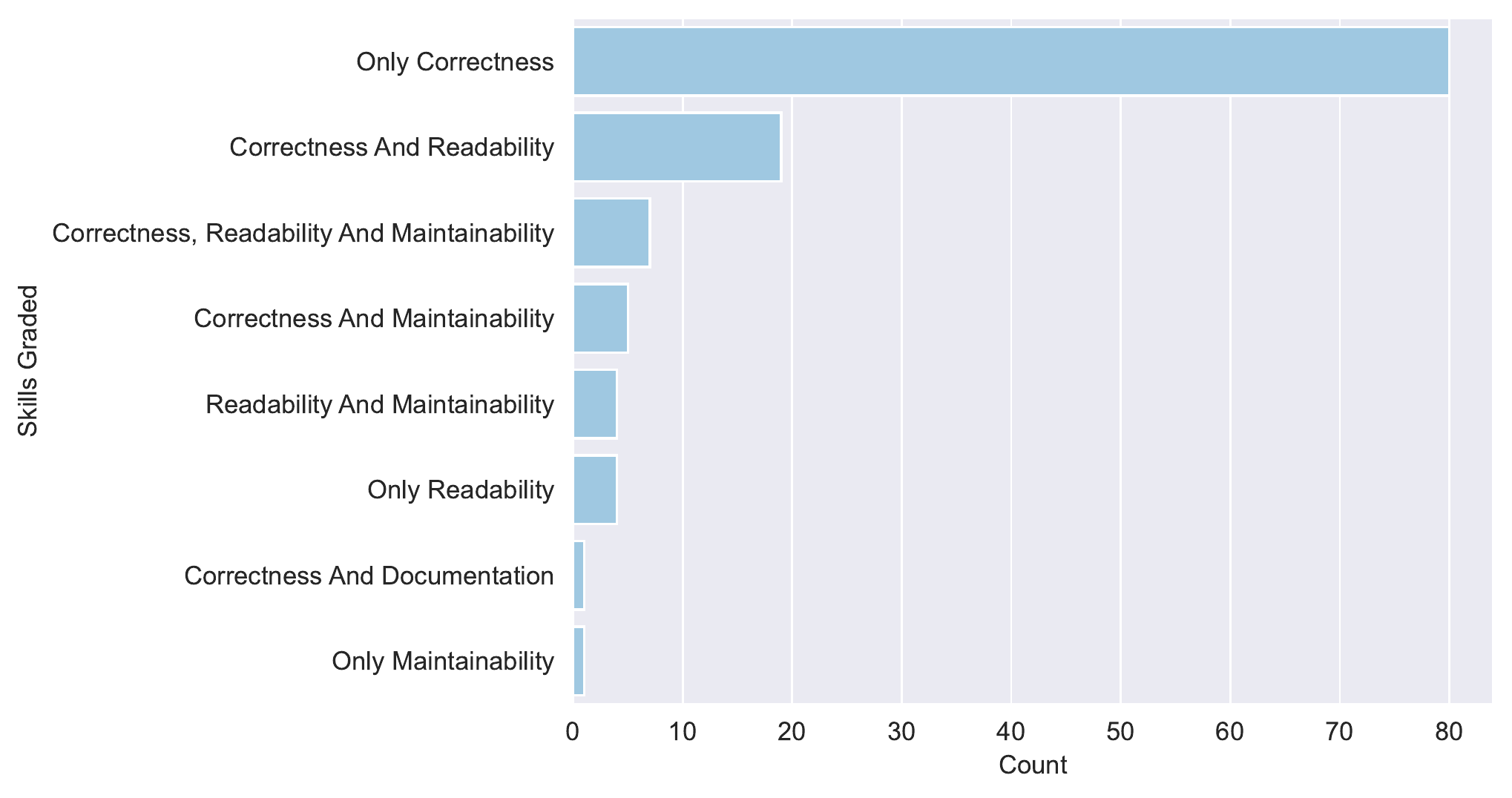}
        \caption{The count of skills assessed, including tools that assess more than one skill.}
        \label{fig:skills}
    \end{figure}

Figure \ref{fig:skills} shows that most of the tools focus on assessing correctness, followed by tools that grade correctness and readability or correctness and maintainability.
Minimal research within our time frame has focused on grading documentation or exclusively maintainability or readability.

\begin{figure}
        \centering
        \includegraphics[width=0.65\linewidth]{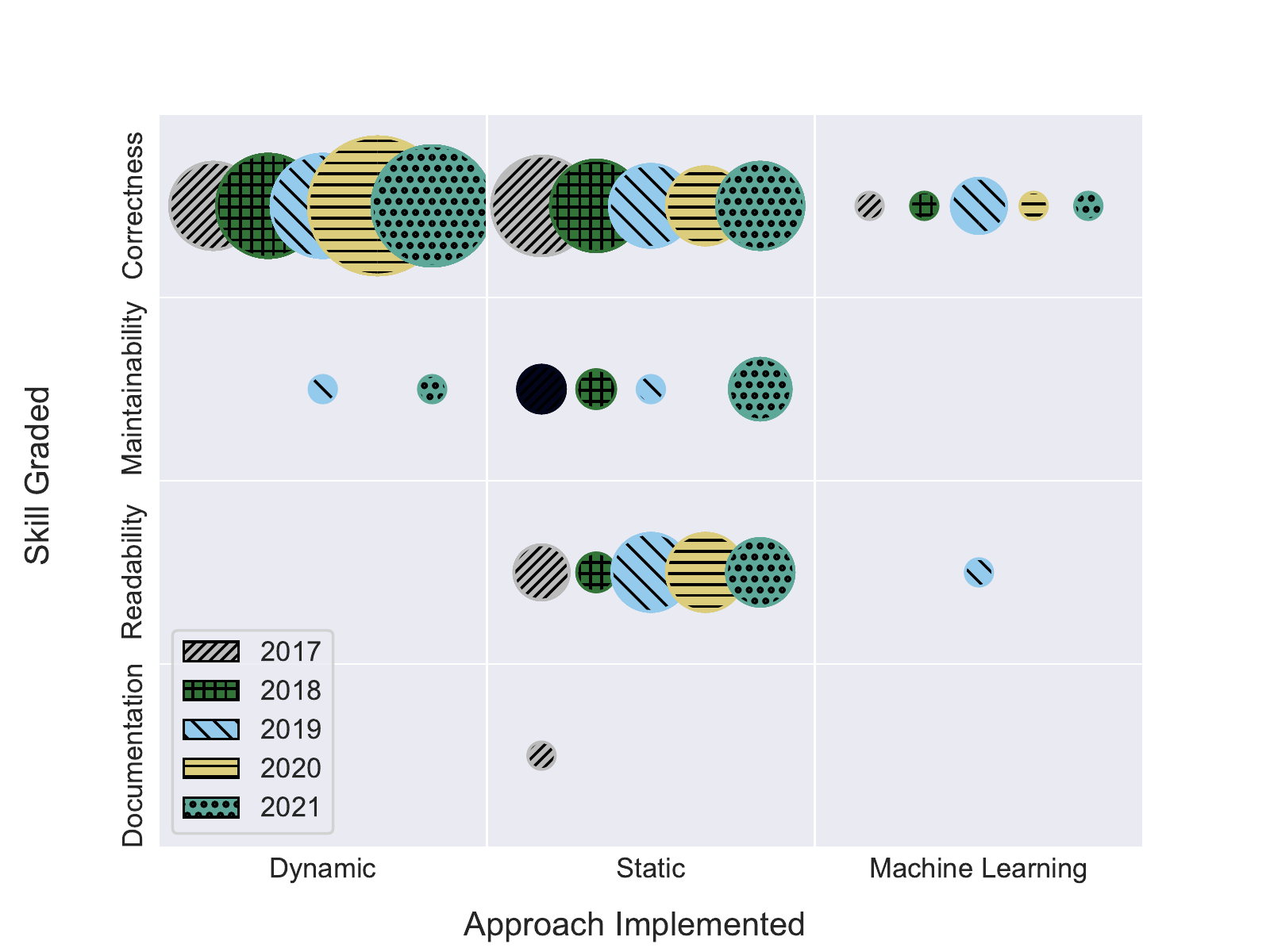}
        \caption{The relative total count of tools categorised by year, assessed skill, and combination of approaches. The size of each bubble is the relative total count of tools categorised, with the larger the bubble the larger the number of tools categorised in research papers published in a particular year.}
        \label{fig:skill_cat_time}
    \end{figure}

Figure \ref{fig:skill_cat_time} shows the combination of skills graded for each category of AAT and year.
Most tools use a dynamic or static approach to assess correctness, and typically static analysis is used to grade readability and maintainability.
Few papers investigate how machine learning can be used to assess any skill.

\begin{figure}
        \centering
       \includegraphics[width=0.85\linewidth]{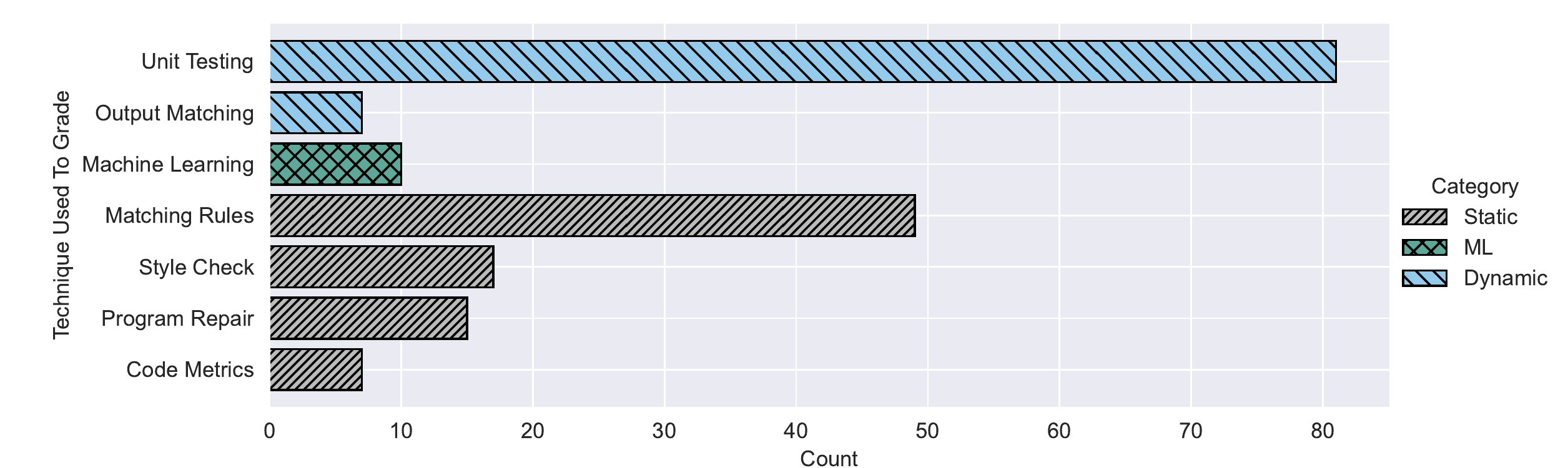}
        \caption{The count of methods classified by the approach utilised to grade and generate feedback, as defined in Section \ref{sect:cats}. ``Matching Rules'' includes techniques such as comparing submissions with model solutions, previous student solutions and use of domain-specific languages. Results with less than five occurrences are omitted.}
        \label{fig:tech}
    \end{figure}

To grade and give feedback on the core skills, different techniques were used depending on the category of the AAT.
Figure \ref{fig:tech} shows the count of techniques used by category, as defined in Section \ref{sect:cats}.
It shows that most dynamic AATs use unit testing, and most other techniques use static approaches.
However, few used techniques such as machine learning or analysing the graphical output \cite{rayyan-354359310}.


\begin{figure}
        \centering
        \includegraphics[width=0.5\linewidth]{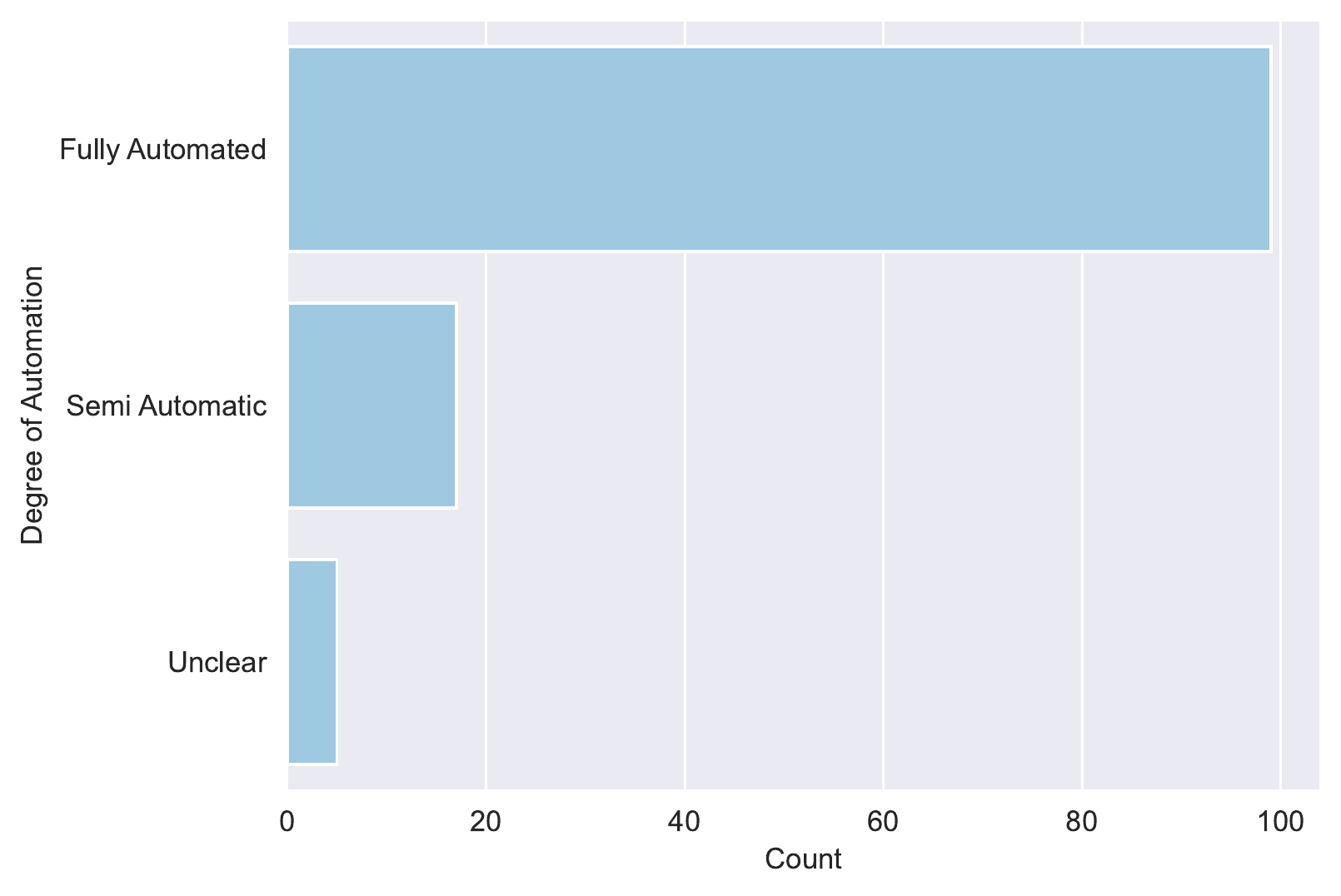}
        \caption{The count of tools that are fully automated or semi-automated.}
        \label{fig:auto_deg}
    \end{figure}

Figure \ref{fig:auto_deg} shows the count of the degree of automation, with most tools implementing fully automated graders.
For a few tools, the publication was unclear on their automation approach.

81\% of tools utilise a fully automated approach to assessment, while others have opted for a semi-automated approach (14\%).
Most AATs are implemented using a fully automated approach to allow for near-instantaneous feedback and multiple resubmissions of assignments without increasing the grading workload.
In contrast, semi-automated approaches are typically used to verify grades given by an AAT or to aid a manual grading or feedback.

Most fully AATs offer bespoke solutions or implement existing grading tools, such as Virtual Programming Lab \cite{Rodriguez2012}, to grade primarily grade correctness, typically by implementing some form of unit tests.
Some tools have adapted continuous integration and delivery by using industry tools such as GitHub Actions\footnote{GitHub Actions -- A commercially available CI/CD tools: \url{https://github.com/features/actions}} and TravisCI\footnote{TravisCI -- Another CI/CD tool: \url{https://www.travis-ci.com/}} to conduct unit tests when students push their code to a repository.
Students typically receive the test results as feedback directly from the chosen tool.

Semi-automated assessment can be implemented using a variety of techniques.
\citet{rayyan-354359291}'s tool has instructor interactions through their grading workflow at multiple stages to grade computer graphics assignments.
Instructors are expected to provide reference solutions, tests, read reports on outliers and highlight source code to produce a rubric, and finally grade the assignments based on the previous test results, extracted rubrics and highlighted code.
Whereas in \citet{rayyan-354359270}'s grader, the instructors' role is to manually grade any assignments that cannot be automatically graded and correct any issues with the automatically generated test cases.

\subsubsection{Correctness}
66\% of the research included in this review only focused on assessing correctness.
Approximately 36\% of AATs focused entirely on correctness and used only dynamic analysis in unit testing.
The others used static analysis (17\%), a combination of dynamic and static analysis (21\%), machine learning in combination with static, dynamic analysis, or as a standalone model (6\%).
Most static analysis tools implement some form of comparison to a model solution or a set of existing solutions to grade or give feedback on a task.
Whereas most dynamic and static analysis tools combine unit testing and comparison to other solutions. 

\citet{rayyan-354359319} developed a gamified web-based automated grader for C programs.
They utilise unit tests for both calculating the grade and providing specific feedback. 
Additionally, they award a certain amount of experience points and a badge for completing the assignment to a certain level.
To evaluate their tool, they conducted a student survey and concluded that it was helpful and that users appreciated the gamification system.

Similarly, \citet{rayyan-354359293} developed a unit test-based tool for automatically grading Android applications.
They employ an industry-standard continuous integration tool to automatically run the test suite when students push to their version control repository.
The students are awarded specific points for each test case passed, with the sum being their total grade.
The authors conducted a student survey and found that most spoke highly of the system, but some students found that the grading logic is not flexible enough, a known issue with unit test-based AATs.

While many automated assessment tools utilise dynamic approaches, some use a static approach.
\citet{rayyan-354359388} developed a grader that uses an instructor-provided reference solution and verifications to grade a submission's correctness.
The verifications allow instructors to adjust the assignment difficulty and associate the verifications with specific language elements.
When evaluating their tool, they found that students' success rate improved, using the framework took less effort from the instructors, and they had a favourable reception from the students.

Dynamic and static approaches are most commonly used to automate the assessment of programming assignments. 
However, some tools have opted to use machine learning to assess the correctness of students' programming assignments.
\citet{rayyan-354359341} trained a Support Vector Machine to grade assignments based on their structural similarity.
They first convert the submissions to their abstract syntax trees (ASTs) and then replace all the identifiers with a standard character.
Using these ASTs, they can calculate the structural similarity and produce a final score.

\citet{rayyan-354359367} and \citet{rayyan-354359295} adapted natural language processing techniques to assess source code.
They normalised the source by removing comments and normalising the names and string literals.
\citeauthor{rayyan-354359367} encode the source code as a binary vector, while \citeauthor{rayyan-354359295} trained a skip-gram model, which is a neural network that predicts a word using the context words around the missing word, to produce their vector encoding.
Finally, they both train neural networks for a final task, \citeauthor{rayyan-354359367} used a dense neural network to predict how to repair an incorrect solution, and \citeauthor{rayyan-354359295} used a convolutional neural network to evaluate the quality of the assignments based on their underlying semantic structure.

\subsubsection{Maintainability}
While most AATs focus on correctness, only one tool in our review assesses exclusively maintainability.
\citet{rayyan-354359378} investigated conceptual feedback vs traditional feedback using a tool called Testing Tutor.
Testing Tutor students learn how to create higher-quality test suites and improve their testing abilities. 
The ability to create a high-quality test suite allows the students to develop higher-quality code by finding bugs in their implementation.
Additionally, creating a suite of test cases allows for future regression testing, tests you can run every time the program changes. These are a vital part of creating maintainable software \cite{kaner1999testing}.

Testing Tutor utilises a reference solution to detect missing test cases or fundamental concepts, such as testing boundary conditions or data integrity. 
This paper gives feedback in either a detailed coverage report or conceptual feedback detailing a core concept that has been missed in the testing.

\subsubsection{Readability}
Four tools focus on assessing only readability and use a static analysis approach for their grading or feedback \citep{rayyan-354359306,rayyan-354359360,rayyan-354359362,rayyan-354359398}.

\citet{rayyan-354359360} implemented a tool to promote learning code quality using automated feedback.
Their tool recommends improvements in a student's code and comments by providing suggestions that the student can implement.

They check that the naming has a consistent style, all names are written in \texttt{camelCase} or \texttt{snake\_case}, check for misspelt subwords and validate if the name contains meaningful subwords.
Meaningful subwords should only contain letters and not contain any stop words.
They implement similar suggestions for comments by checking that comments do not have any misspelt words, meaningless words, or comments that are too short.
To evaluate the tool's effectiveness, they conducted two experiments. 
The first was conducted using a control group and an intervention group, and the second focused on a year one and a year two programming course.
They conclude that while the research is incomplete, the tool can be helpful, as students do not satisfy all the code quality requirements due to human error.

Similarly, \citet{rayyan-354359398} extended an existing Python static analysis by developing custom checks and feedback for novice programmers.
They provide a textual description of the error, provide an example, and explain why the error is problematic.
To evaluate their tool, they compared two years of an introductory Python course, with one year using the tool and the previous year without.
The year that utilised the authors' tool significantly reduced the number of repeated errors per submission and the number of submissions required to pass the exercise.

\subsubsection{Documentation}
No papers focus entirely on documentation.
However, \citet{rayyan-354359305} introduced AppGrader, an automated grader to grade Visual Basic applications, which graded correctness alongside documentation.
They used static analysis to check if best practices have been followed and if comments exist for each subroutine or function.
However, the tool did not analyse the quality of the documentation.
In their bench tests, they assessed the tool against two scenarios, a typical homework assignment for an introductory programming course and the other was the source code for the tool itself, to find the overall execution times.
The average overall execution time was 8.54 seconds for the typical home assignments, allowing this tool to be used as near-instantaneous feedback during development.

\subsubsection{Combination Graders}
While some AATs only focus on grading one skill, others aim to grade a combination of skills with varying approaches.
The most common is to assess readability alongside correctness (15\%).
These tools primarily use a dynamic approach to assess correctness and a static approach to assess readability.

Some tools assess maintainability in addition to correctness and readability (6\%) or just in addition to correctness (4\%). 
These typically use either dynamic approaches in the form of mutation tests to assess the quality of the student-created test suites or static approaches such as metrics to calculate the complexity of the code.
Few tools focus exclusively on assessing maintainability and readability (3\%).
These AATs use a static approach to assess both these skills.

\citet{rayyan-354359369} introduced Annete, an intelligent tutor for Eclipse, to give feedback on correctness and readability.
They use a neural network with a supervised learning algorithm to determine if a student needs assistance with their code and determines what feedback should be shown.
Annete can give multiple types of feedback, including feedback about how to proceed, such as using language structures that should not be used in a particular assignment and practical support, such as giving positive feedback when they have passed a test case.

Similarly, \citet{rayyan-354359351} developed a tool that provides feedback on maintainability and readability during development and then used unit tests to grade correctness after the final submission.
For the feedback on maintainability and readability, they utilised static analysis to find \textit{antipatterns}, including commonly occurring practices that reflect misunderstanding, poor design choices and code style violations in students' submissions.

While most tools provide exclusively textual feedback, \citet{rayyan-354359333} have developed a tool that combines spectrum-based fault localisation with visualisation to provide feedback on maintainability and uses unit testing to grade and provide feedback on correctness.
They visualise the spectrum analysis as a heatmap, with more suspicious code producing a higher score.
To evaluate their tool, they conducted a user study over two semesters. 
One was a control group that received only textual feedback, and the other semester provided both textual feedback and heat map visualisation.
They found that having access to heat maps made it easier for students to make more incremental progress towards maximising their scores.

\subsection{Languages Evaluated (RQ2)}
\begin{figure}[h!t]
    \centering
    \includegraphics[width=0.65\textwidth]{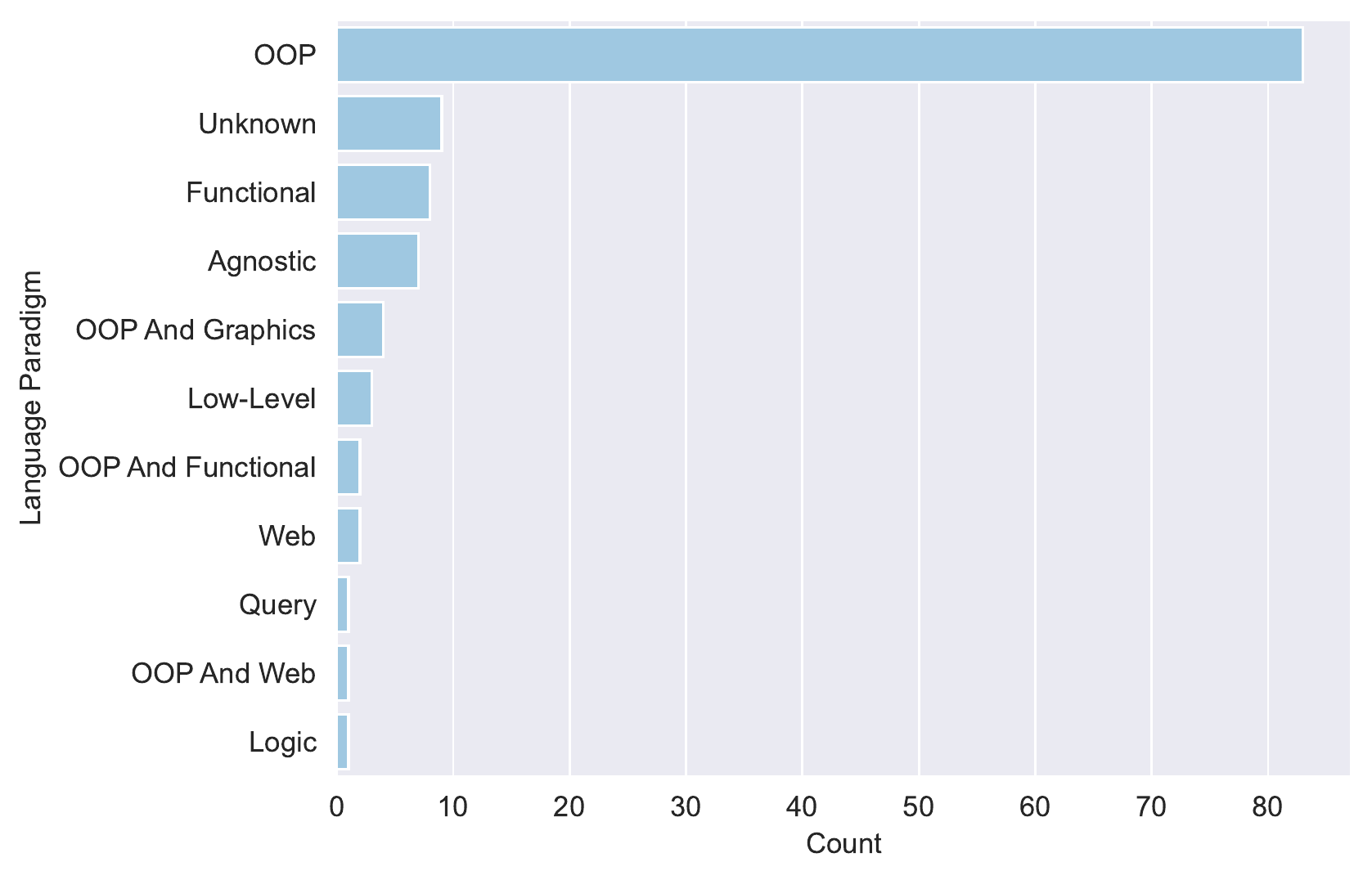}
    \caption{The count of the language paradigms assessed, including tools that grade more than one language paradigm. ``OOP and Graphics'' are tools that assess the development of shaders and graphics using C/C++, and ``Agnostic'' includes tools that specify that any language can utilise the tool.}
    \label{fig:lang}
\end{figure}

As AATs can focus on grading any number of languages, we grouped the languages into their primary paradigm.
Figure \ref{fig:lang} shows the count of language paradigm among the tools.
Most tools assess OOP languages, followed by functional languages and tools designed to grade any language\footnote{Non-OO or Imperative languages did not solely appear in the literature, and were paired with OO languages. Languages that can be used in an OO or non-OO way are included in the OO category.}. 
Some papers did not clearly specify which language the tool assesses; these have been annotated as unknown.

\subsubsection{Object-Oriented, Functional and Logical Programming Languages}
Most AATs grade object-oriented languages (69\%), such as Java and Python.
Due to object-oriented languages' prominence in education and industry, object-oriented graders are the most developed, making practical teaching tools for the programming approaches we used \cite{Kolling1999}. 

\citet{rayyan-354359410} developed a tool to generate personalised hints for Python.
Their tool first finds successful submissions that have failed the same test at some point and then computes an edit path from the student's incorrect submission to the successful submissions.
They then translate the edit path to natural language hints.
To evaluate their tool, they used a set of historical submissions and measured the percentage of submissions for which the tool can generate hints.
They found that their tool could generate a personalised hint for most submissions while only using data from as little as ten previous submissions.

While \citeauthor{rayyan-354359410} focused on generating personalised hints for Python assignments, other authors focused on automated grading of Java.
\citet{rayyan-354359383} developed a tool to aid teaching assistants when manually reviewing submissions by generating a report assessing how well the student performed against a set of learning objectives.
Similarly, \citet{rayyan-354359337}'s tool aimed to aid graders with grading the correctness of methodology; explicitly do students' submissions contain the required methods or fields rather than the correctness of the functionality.

\citet{rayyan-354359347} utilised formal semantics to develop a real-time Python AAT.
They use a single reference solution to find differences in the output and execution trace of a student's submission. 
The student's submission is graded as incorrect if a difference is found.
They evaluated their tool against a set of benchmarks of existing submissions graded by test suites and discovered that it revealed no false negatives, but a test suite did since it was missing some test cases.

Similarly, \citet{rayyan-354359325} developed a tool that uses semantic analysis, a knowledge base of programming patterns and the instructor's input to correlate the patterns with detailed natural language feedback to provide personalised feedback for Java assignments.
They compared their work against state-of-the-art techniques, including \citeauthor{rayyan-354359347}'s tool, and concluded that their approach is based on understanding the semantics of the submissions and the original intentions of students when dealing with an assignment.

In addition to assessing object-oriented languages, some AATs assess functional languages, typically Haskell or OCaml \citep{rayyan-354359307,rayyan-354359314,rayyan-354359327,rayyan-354359336,rayyan-354359354,rayyan-354359375,rayyan-354359397,rayyan-354359401,rayyan-354359403,rayyan-354359412}.
\citet{rayyan-354359336} developed an online IDE and AAT for OCaml while designing a massive open online course (MOOC). 
The IDE provided syntax and type error feedback as annotations and graded student submissions using unit tests.

While \citeauthor{rayyan-354359336} developed an IDE as part of a MOOC, \citet{rayyan-354359401} developed a program repair-based feedback tool for OCaml.
They utilised test cases and correct reference solutions to find the fault and repair logical errors in students' submissions. 
They conducted a benchmark and student survey to evaluate the tool's efficiency and helpfulness. 
They found that the tool is powerful and capable of fixing logical errors in student submissions and that the students found it helpful.

\citet{rayyan-354359375} developed an alternative program repair-based feedback tool for OCaml, utilising multiple partially matching solutions and test cases to generate a fix for logical errors.
They compared their tool to \citeauthor{rayyan-354359401}'s tool and found that their approach was more effective at repairing submissions.
Additionally, they conducted a user study, and students agreed that their tool was helpful.

Other tools have generated ``am I on the right track'' feedback for other functional languages.
\citet{rayyan-354359397} conducted a pilot study of a tool for Scheme that transformed a student's partial submission into a final program with the same functionality as a desired correct solution to determine if the student is on the right track.

Similarly, \citet{rayyan-354359412} developed a tool for Haskell, which used programming strategies derived from instructor-annotated model solutions to determine if a student's submission is equivalent to a model solution.
They deployed their tool into their course, and most interactions with the tool were classified as correct or incorrect.
Furthermore, they conducted a student survey to evaluate the perceived usefulness of the tool. 
They found that students were taking larger steps than the tool could handle, that the tool was sufficient, and that there was room for improvement. 

Only one paper grades a logic language: \citet{rayyan-354359414} has developed a tool to assess Prolog clauses using static analysis.
To grade and give feedback on Prolog clauses, the authors convert the clauses to abstract syntax trees and extract patterns that encode relations between nodes and the program's syntax tree. 
These abstract syntax tree patterns are then used to predict program correctness and generate hints based on missing or incorrect patterns.
They evaluated their approach on past student assignments and found that the tool helps classify Prolog programs and can be used to provide valuable hints for incorrect submissions.
However, more work must be done to make the hints more understandable by annotating natural language explanations of their patterns and derived rules.

\subsubsection{Specific Language Domains}
While some AATs focus on the three major language paradigms, other AATs focus on grading more specific areas, such as web-based languages \citep{rayyan-354359298,rayyan-354359362,rayyan-354359386}, graphics development \citep{rayyan-354359280,rayyan-354359291,rayyan-354359340,rayyan-354359349}, kernel development and assembly languages \citep{rayyan-354359366,rayyan-354359373, rayyan-354359390}, or query languages \cite{rayyan-354359278}.

\citet{rayyan-354359362} introduced a tool to grade the quality of web-based team projects and measure students' contributions. 
The authors utilise continuous integration tools to analyze the version control logs to determine the students' contribution and use existing static analysis tools, including SonarQube and StyleLint, to evaluate the quality of the source code.
While applying this tool to a course in their department, they found a weak association between lines of code modified and the final grade.
Students with better grades fixed more errors but also introduced more errors.

To assess computer graphics courses, \citet{rayyan-354359340} and \citet{rayyan-354359349} developed AATs to grade OpenGL.
Both solutions compared the students' output to a reference solution by comparing the difference in pixels, with \citeauthor{rayyan-354359349} also grading based on parameters passed to OpenGL and algorithm results, such as outputs of parametric equations.
\citeauthor{rayyan-354359340} implemented ``visual unit testing'', allowing instructors to script keyboard and mouse input and provide detailed feedback through screenshots and videos of the execution.
They both conducted student surveys to evaluate their tools' usefulness for learning computer graphics and reported that the students found the tools helpful when learning to program computer graphics.

To assess operating system kernels \citet{rayyan-354359390} developed a cloud-based Linux kernel practice environment and judgement system.
This uses a dynamic approach to test the students' attempts at programming an operating system kernel by comparing their kernel outputs to the teacher's reference solution.
The authors evaluated the parallelisation of their tool by measuring the execution time when running tests and comparing these results to a baseline serial execution, with their tool taking less time to grade longer scripts.
Finally, they concluded that their tool works well in the real world and reduces the effort to verify a student's work.

While \citeauthor{rayyan-354359390} focused on assessing kernel-level assignments, \citet{rayyan-354359373} and \citet{rayyan-354359366} developed tools to assess assembly code assignments.
Both tools utilised a dynamic approach in the form of test cases to grade and give feedback on their assignments.
Additionally, \citet{rayyan-354359366} also implemented an IDE plugin to give continuous feedback to students, including errors that the code cannot assemble, warnings to indicate potential bugs and information based on the analysis of the code.
To evaluate their tools, they surveyed students and found that their tools were beneficial and contributed positively to the student's learning of assembly languages.

To grade SQL statements, \citet{rayyan-354359278} investigate combining dynamic and static analysis.
The dynamic analysis compares the output of the students' statements with the expected value.
The static analysis compares the syntax similarity using an abstract syntax tree and the textual similarity of the statements themselves.
To evaluate their approach, they compared their hybrid approach to a dynamic analysis that executes and compares the results with the expected results, a syntax-based approach that calculates the syntax similarity between a submission and a reference solution using the abstract syntax tree and a text-based approach that calculates the textual similarity between the student's statement and a reference statement.
They found that the existing grading approaches could not yield satisfactory results and that the hybrid approach successfully identifies various correct statements submitted by students and grades other statements according to their similarities.

\subsection{Techniques Used to Evaluate the Tools (RQ4, RQ5)}\label{sect:results:eval}
\begin{figure}[b]
    \centering
    \includegraphics[width=0.55\linewidth]{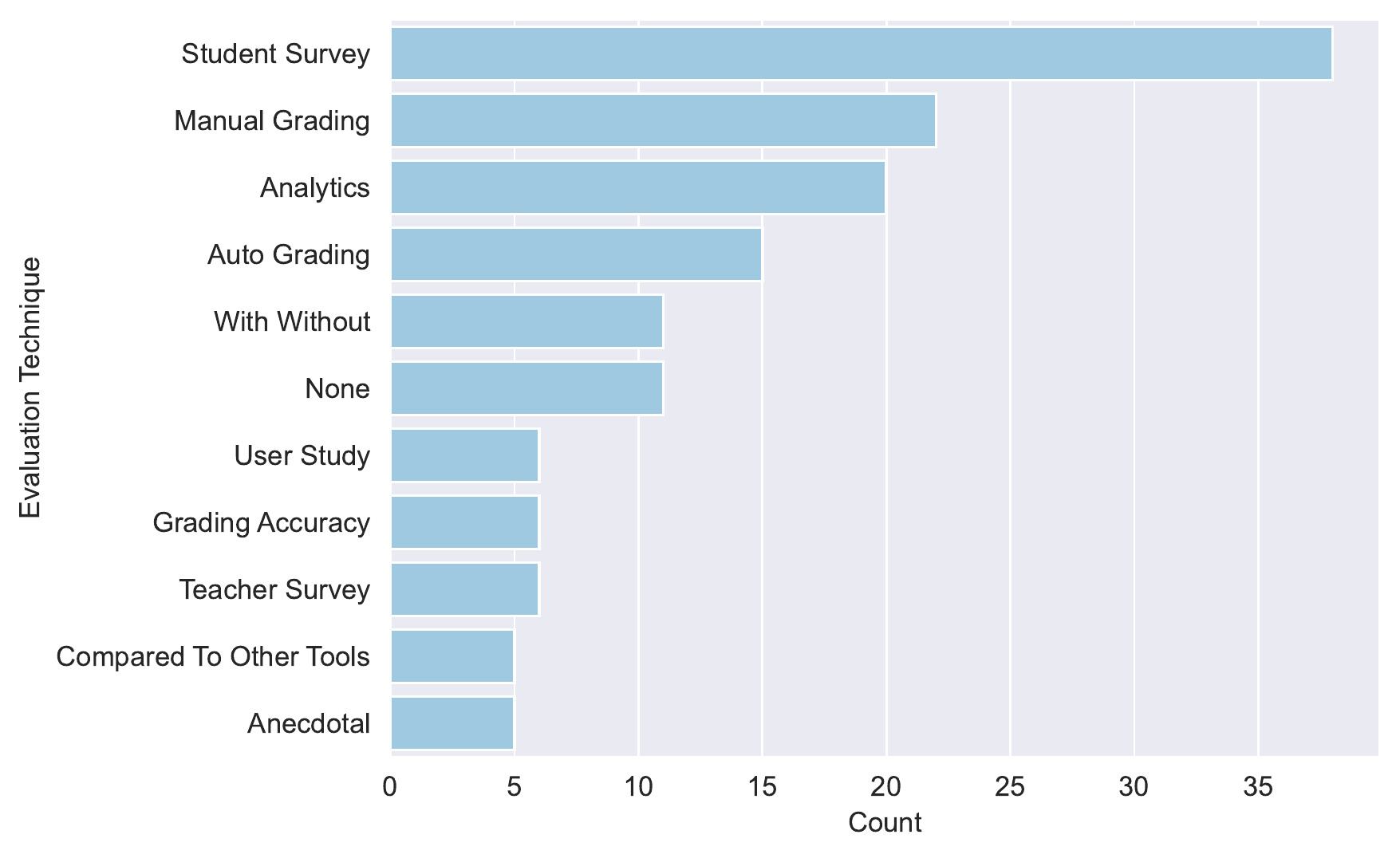}
    \caption{The count of papers implementing these techniques to evaluate the automated grades. Results with under five occurrences are omitted.}
    \label{fig:eval}
\end{figure}

\begin{figure}
    \centering
    \includegraphics[width=0.5\linewidth]{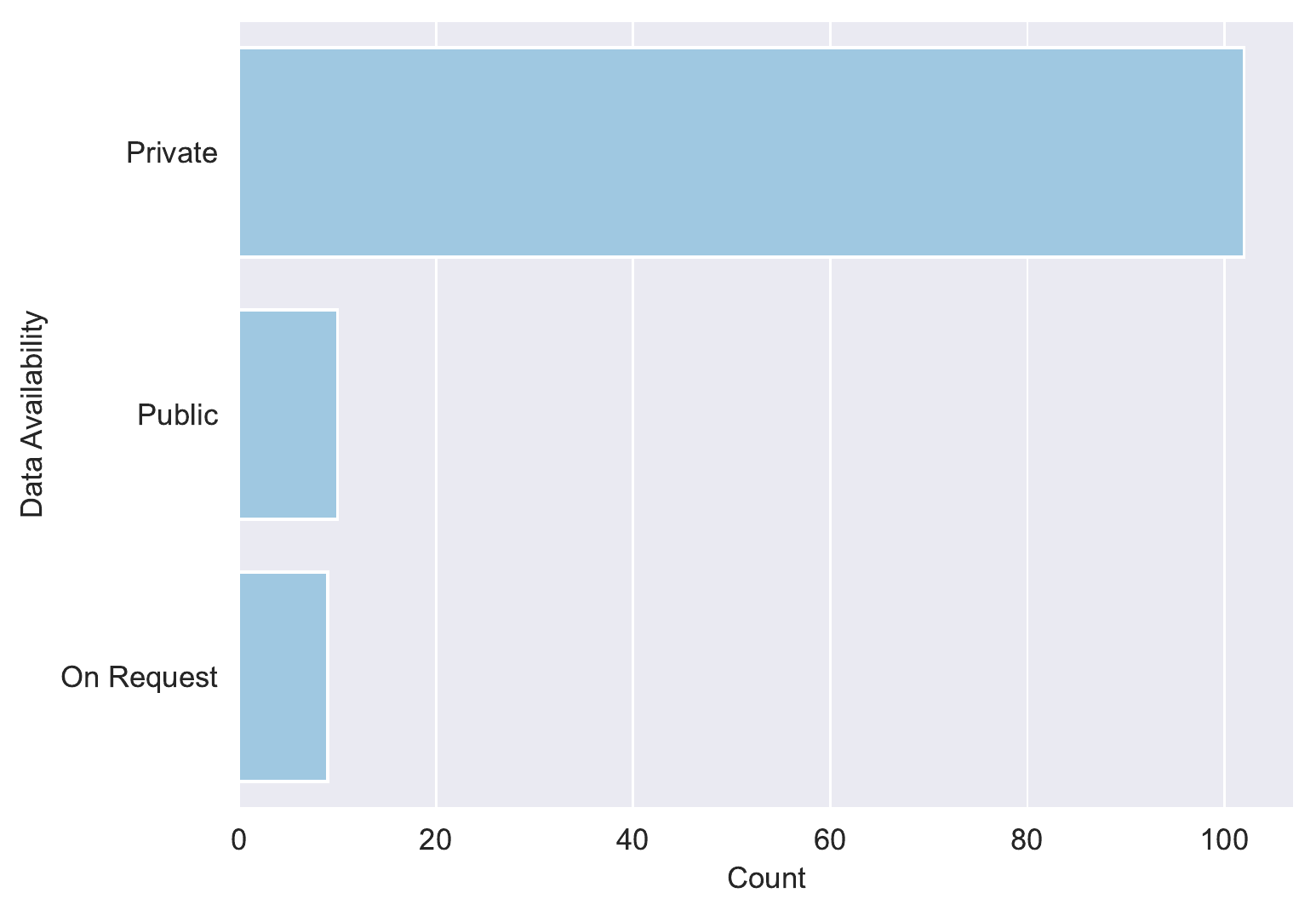}
    \caption{The count of data availability used in the evaluation.}
    \label{fig:data_ava}
    \vspace{-1.8em}
\end{figure}

Experiments were conducted using different techniques to evaluate the tool's quality.
Figure \ref{fig:eval} shows the count of techniques used in the papers, with student surveys and tools being compared to manual grading being the most common.
Most of these experiments were conducted on course-specific assignments (66\%). 
As such Figure \ref{fig:data_ava} shows that most of the data is not available to validate the results or for future research.
While most tools are evaluated in some form, most do not provide the dataset in which the evaluation occurs (84\%); this might be due to most tools being evaluated against course-specific assignments (66\%) or exams (10\%), which typically are not be distributed publicly.


\subsubsection{Approaches}
90\% of the papers include some form of evaluation, with most evaluations conducted by the tool developer.
The primary method of evaluation is to ask for student feedback on a tool in the form of surveys (24\%), as they are the primary beneficiaries of AATs.
Their insight provides valuable feedback on how well a tool helps them learn how to program.

Another typical evaluation approach compares the tool to either manual grading (22\%) or other automated tools (9\%).
The evaluation is typically used to check the tool's grade accuracy compared to manual grading.
Other aspects, such as using an automated tool to improve students' grades or allow them to receive feedback faster, are also evaluated.
Though most of the included articles perform some form of evaluation, 10\% of them do not; this could be due to the tools still being in development and not at a stage where a full evaluation would be beneficial.
However, even an initial validation that the tool is performing as expected at the early stages of the project could be beneficial to confirm that the final tool will aid students' learning.

\subsubsection{Performance}
As an exploratory analysis, we annotated a third of all papers that conducted some form of evaluation and how well they performed.
We found that most tools perform well in their evaluations, though some authors report mixed results, especially if using multiple evaluation approaches.
While the tool developers conducted most evaluations, some tools were evaluated by third parties, typically in papers evaluating multiple tools or papers reporting authors' experiences using existing tools.
Reporting only positive results makes it difficult to evaluate different tools reliably and can be challenging for instructors to select the most effective tool for their course, especially with most AATs implementing the same methodology.

Some authors have produced evaluation or experience reports discussing one or more tools to provide instructors with an external evaluation of some AATs.
\citet{rayyan-354359364} evaluated two AATs, one that gave personalised hints using program repair and a program visualisation tool, to determine if students can solve problems faster, understand problem-solving and fix bugs easier, compared to a test suite.
In their experiment, they had three test conditions; in the base condition, students only had access to the test suite, and in the other two, they had access to one of the tools and the test suite.
The students were asked to complete a set of problems using one of the three conditions randomly assigned to the problem; after reaching a correct solution, they were presented with a post-test.
This post-test was used to evaluate the student's understanding of how to solve the problem and consisted of students being presented with four different solutions for a problem and being asked to identify which solutions were correct or incorrect without being able to execute the program.
They found that the program repair tool greatly cuts student effort, with fewer attempts, and students using the visualisation tool showed lower post-test performance.

Similarly, \citet{rayyan-354359309} evaluated two AATs, one implemented using unit tests and the other utilising reference solutions.
They focused on applying these AATs in the context of a massive open online course. 
They found that the reference solution-based AAT typically performs as well as the unit test-based approach.
However, the reference solution-based tool awards lower grades to correct solutions that are rarely implemented.

While some evaluation papers compare existing tools, \citet{rayyan-354359316} investigate how different test suites can provide different grades and how the properties of the unit tests impact the awarded grades.
To answer their research questions, they extended an existing set of programming assignments with artificial faulty versions and a sample of test suites from a larger pool.
They generated the grades by calculating the percentage of passed tests and compared the different test suites.
The authors concluded that the grades vary significantly across different test suites and that code coverage, the percentage of the source code executed while running a test suite, affects generated grades the most.

\subsection{Performance Against Human Graders (RQ4, RQ5)}
Only 22 tools are evaluated against human assessors, with 16 focused on grading and six focused on feedback.
While some of the automated graders provide feedback to students, typically in the form of unit test results, similarity to model solutions or predefined human feedback, human assessors have not evaluated the generated feedback. 
To investigate how well AATs perform when compared to humans, we further annotated papers that conducted an evaluation that included some comparison to the assessment provided by a human.

\begin{figure}
    \centering
    \includegraphics[width=0.85\linewidth]{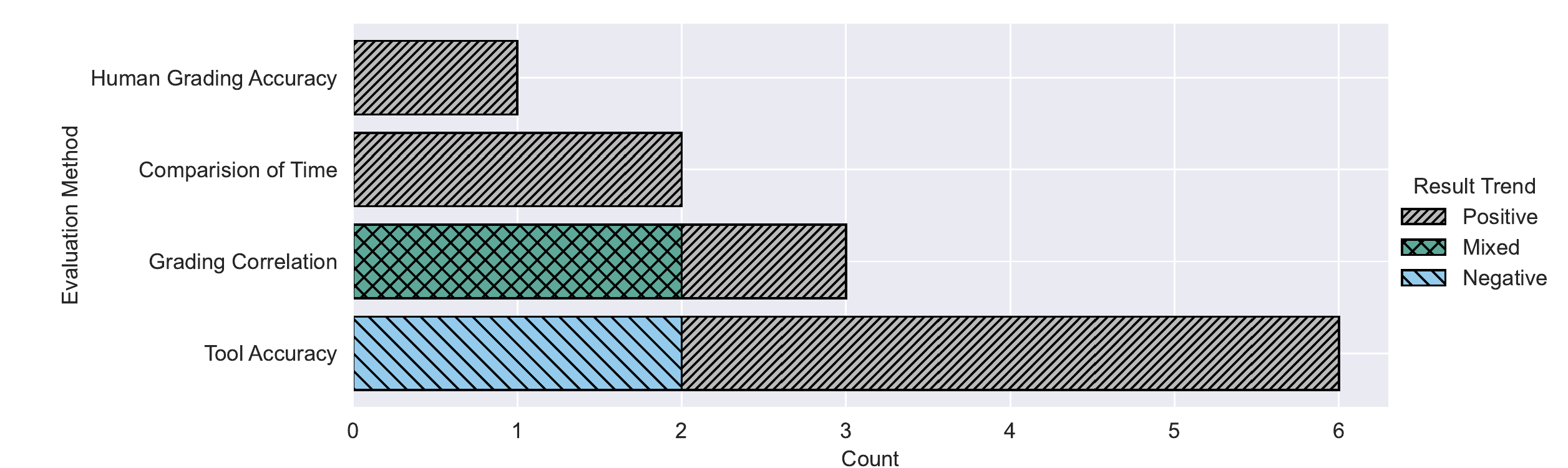}
    \caption{The count of automated graders categorised by the evaluation technique when compared to human graders. The resulting trend is the authors' sentiment of the performance when compared to humans. In the case of human grading accuracy, the AAT was taken as the ground truth and the human graders were found to be more error-prone \cite{rayyan-354359297}.}
    \label{fig:grading_compared_to_human}
\end{figure}

Figure \ref{fig:grading_compared_to_human} shows the count of tools by different evaluation techniques that involved comparing the results from the AAT with a human and the authors' sentiment of the performance of the AAT.
Most are evaluated against the accuracy of the AAT compared to human-provided grades \cite{rayyan-354359274, rayyan-354359278, rayyan-354359279, rayyan-354359302, rayyan-354359321, rayyan-354359323, rayyan-354359332, rayyan-354359341}, with most reporting positive results.

\citet{rayyan-354359278} tool uses a hybrid approach of reference solutions to automate the grading of SQL statements. 
It is evaluated against a benchmark of human-graded submissions and three other state-of-the-art approaches, including dynamic, syntax-based and text-based analysis. 
The proposed approach performs better than other state-of-the-art approaches with a clear advantage, with a mean average error of 8.37, compared to 26.01 for static analysis, 29.89 for text-based analysis and 21.66 for dynamic analysis.

While \citet{rayyan-354359279} primarily discusses automated feedback in the form of code repair suggestions, the evaluation compared to a human is focused solely on the effect the code repair has on the automated grading and not how the feedback compares to a human assessor.
The evaluation shows that their code repair tool increases the grade precision, measured by comparing the submission's auto-graded result before and after code repair to manually graded scores, of their unit test-based automated grade, by 4\%, from 81\% to 85\%.
This shows that their tool improves the precision of the automated grader by repairing uncompilable code that can then be automatically graded.

Most tools that evaluate the accuracy of automated graders take the human-provided grades as ground truth, \citet{rayyan-354359297}, introduce JavAssess, a framework that automatically inspects, tests, marks and corrects Java source code. 
To evaluate the performance of JavAssess to human graders, they compare the accuracy of manually and automatically graded exams. 
However, they treat the automatically provided grades as the ground truth and conduct an ANOVA test to validate the influence of human graders on marking errors. 
They concluded that human graders influence the mark when marking, with the probability value associated with the F value $Pr(> F) = 0.0164$ below the significance level of 0.05. 

While most papers evaluate the tools' accuracy, some investigate the correlation between human-provided grades and automated tools \cite{rayyan-354359272, rayyan-354359294, rayyan-354359326, rayyan-354359300, rayyan-354359334}.
\citet{rayyan-354359272} utilise unit testing and pattern rules to automate the grading of six programming assignments within a single undergraduate course.
They compared the human grades to the tool's grades to evaluate their tool. 
They found that the tool's grades were significantly positively correlated with human grading in all six assignments $(p < 0.001)$, with on average 70\% of the tool's grades being the same as the human graders.

\citet{rayyan-354359300} provide a detailed statistical analysis comparing manual and automated assessment of programming assignments and report on a lecturer's experience integrating automated assessment into their module.
To evaluate their chosen AAT, they used 77 exams that were manually and automatically graded and were analysed using a paired T-test, and the correlation between manual and automated assessment was measured with Pearson's correlation coefficient. 

Finally, the students were classified into three categories (failing, passing, and passing with distinction) based on their manual assessment marks, and a T-test was then performed on the categorical data. 
The T-test resulted in a ``medium practically visible difference found'' and can be attributed to the difference in granularity between the marks given by manual assessment and those given by unit tests and the correlation proved to be a significant relationship $(r = 0.789, p < 0.001)$. 
The t-test results on the categorical data show a very large, practically significant difference for failing students $(d = 1.303, p < 0.001)$ and students who passed $(d = 0.151, p = 0.409)$, suggesting that automatic assessment is not reliable. 
However, they found that automated assessment can be more trustworthy for higher achieving students, with the small effect size $(d = 0.151, p = 0.409)$.

Another factor in automating grading is reducing the time instructors take to grade. 
Two papers investigate the effect of the time taken to grade assignments, with and without their tool \cite{rayyan-354359270, rayyan-354359310}.
\citet{rayyan-354359270} introduce a semi-automated approach to grading Java assignments by automatically generating unit tests from an instructor's solution and only presenting the graders with submissions that cannot be graded using the unit tests.
To evaluate how well their tool expedites the grading process, they asked graders to grade six Java exams manually and then, six months later, asked the same six examiners to mark the exams using the tool.
While instructors invested an additional hour, on average, to prepare the exams, the automated tool reduced the average time taken to grade each submission from 6 minutes to 2.5 minutes.
The total average time invested, including preparation, decreased by 25.2 hours, from 53.5 to 22.3 hours.

\begin{table}[b]
\footnotesize
\begin{tabular}{|p{0.2\columnwidth}|l|p{0.17\columnwidth}|p{0.12\columnwidth}|p{0.12\columnwidth}|p{0.12\columnwidth}|p{0.12\columnwidth}|}
\hline
\multicolumn{2}{|l|}{\multirow{2}{*}{}} & \multicolumn{5}{c|}{\textbf{Evaluation Technique}} \\ \cline{3-7} 
\multicolumn{2}{|l|}{} & \textbf{Accuracy of \newline Categorisation} & \textbf{Error \newline Detection} & \textbf{Expert \newline Agreement} & \textbf{Grading \newline Correlation} & \textbf{Time to \newline Bug Fix} \\ \hline
\multicolumn{1}{|l|}{\multirow{2}{*}{\rotatebox[origin=r]{90}{\parbox{0.75cm}{\raggedright \textbf{Result \newline Trend}}}}} & \textbf{Positive} & \cite{rayyan-354359407} & \cite{rayyan-354359405} & \cite{rayyan-354359383} &  &  \\ \cline{2-7} 
\multicolumn{1}{|l|}{} & \textbf{Mixed} & & & \cite{rayyan-354359397} & \cite{rayyan-354359345} & \cite{rayyan-354359367} \newline \\ \hline
\end{tabular}
 \caption{The automated feedback tools are categorised by the evaluation technique when compared to human graders. The resulting trend is the authors' sentiment of the performance when compared to humans.}
\label{tab:feedback_compared_to_human}
\end{table}

Only six papers evaluate the feedback by comparing the automatically generated feedback to human-generated feedback \cite{rayyan-354359367, rayyan-354359397, rayyan-354359407, rayyan-354359345, rayyan-354359383, rayyan-354359405}.
Table \ref{tab:feedback_compared_to_human} shows the count of tools by different evaluation techniques that involved comparing the results from the AAT with a human and the authors' sentiment of the performance of the AAT. 

\citet{rayyan-354359367} use code repair to provide targeted examples for compilation errors.
They evaluate their AAT by comparing the time taken to repair compilation errors with and without the automated code repair feedback, with both groups having access to human teaching assistants (TAs).
They found that the tool helped resolve errors 25\% faster on average, and the large-scale controlled nature of the empirical evaluation implies that the tools are comparable to human TAs.

\citet{rayyan-354359397} utilises instructor-provided solutions to generate "am I on the right track" feedback.
As part of a pilot study, the tool's feedback was shown to a subset of experienced TAs, and all TAs were asked to replicate what they would do when interacting with incorrect student code, and the TAs were asked to provide feedback on the tool.
The TAs in the study provided mixed feedback and suggested areas of improvement, with one TA saying that the fully automated feedback features caused them to have a lack of control over the process.
Other TAs saw the potential for the tool and how "am I on the right track" style feedback can aid struggling students when they do not have access to teaching staff.

\citet{rayyan-354359405} present a virtual teaching assistant to help teachers detect object-oriented errors in students' source code by converting source code to Prolog and inferring errors from instructor-provided rules.
To validate the corrections made by their tool, they used the virtual teaching assistant to check students' coursework previously checked by human graders, to check if the tool overlooks corrections made by the instructors and to reduce the instructors' workload.
The authors observe that the tool detected 125 additional errors, totalling 196 instead of 71 object-oriented errors.
However, they also found three types of object-oriented errors that the tool cannot detect, including non-static methods that do not use any field or method of the object, non-abstracted fields in sister classes and using Java Collections methods without implementing the \texttt{equals} methods. 
They repeated this study with additional rules to handle these missing errors, and the tool found 29 of these specific three errors overlooked by the instructor.

\section{Discussion}\label{sect:discussion}
\subsection{Why the Focus on Correctness Assessment? (RQ3)}
There are many reasons why researchers choose to focus on assessing correctness over other important skills.
Assessing the correctness of functionality teaches students how to analyse and implement features based on written requirements.
Understanding how the requirements can be translated to the desired features is a crucial skill and one that is often used in the industry.
Additionally, assessing the correctness of methodology allows instructors to verify that students understand particular programming concepts, such as conditionals, iteration, and recursion, or other course material, such as particular algorithms they have been asked to implement.

Primarily, assessing correctness uses a dynamic approach in the form of unit tests, which have several benefits and limitations.
Unit tests allow for quickly assessing large quantities of assignments and enable students to receive near-instantaneous feedback from the test suite results.
Implementing a test suite to assess an assignment requires specific unit testing knowledge, and more complex implementations require reflection knowledge.
However, many unit test-based tools aid instructors in creating their test suites by providing a framework to implement them or simplifying unit testing into a set of input-output tests.

While these AATs offer near-instantaneous feedback, they are often simplistic, typically displaying if the test passed or failed, and if it failed, the difference between the expected and actual outputs or any errors.
This limited feedback only informs the student of an issue in their code, either that an error is produced or that their output does not match the desired output, and does not help them resolve their issues.
Providing this form of limited feedback, where the student knows the error and potential location but does not provide any hints on how to fix their issues, is similar to the limited and cryptic feedback of compiler messages, which often results in increased frustration and hampers progress \cite{Becker2016}.

Furthermore, most of these AATs cannot award partial grades for incomplete or uncompilable programs or distinguish between qualitatively different incorrect solutions, resulting in students receiving a zero grade. 
In contrast, if a human graded them manually, they would typically receive partial marks for source code that implements a subset of the features or has minor logical or syntax issues.
Some AATs resolve this by implementing code repair to fix the broken code before running the test suite, thus providing partial grades for syntactically incorrect submissions, typically by deducting marks from the unit tests results of the repaired code \citep{rayyan-354359270,rayyan-354359271,rayyan-354359279,rayyan-354359290,rayyan-354359299,rayyan-354359303}. %

Another limitation of using unit testing to assess correctness is that unit tests typically require students to follow a strict pre-defined structure for their code, often with class and function names specified.
Furthermore, the popularity of unit test-based AATs can influence instructors when designing their assignments.
Suppose the instructors have large class sizes or want to give students near-instantaneous feedback, in that case, they often use a unit test-based AAT, typically leading to small-scale closed-ended assignments.
These small-scale closed-ended assignments and the strict structure can limit students' opportunities to be creative.
Additionally, this strict structure and small scale of the assignments can limit the student's ability to learn how to write maintainable and readable code by limiting their opportunities to learn how to name classes and functions and how to design object-oriented solutions.

Instructors can design open-ended larger-scale coursework to enable students to take responsibility for their assignments and use their creativity to implement the required features.
Students who assume control over their learning experience gain knowledge more effectively than those who do not \cite{Knowles1975}.
Furthermore, allowing students to take ownership of their projects and use their creativity can increase their motivation and learning opportunities \citep{Sharmin2021}.

However, these are often impractical to assess at scale, as assessing open-ended assignments is time-consuming and challenging to automate with traditional auto-grading methods.
A common approach to assessing open-ended assignments is to use multiple human graders \cite{Aziz2015}. 
However, using multiple graders can lead to grade and feedback consistency issues, as readability, maintainability, and documentation are often subjective.
While multiple graders reduce the workload for a single grader, the overall assessment process is still time-consuming.
Additionally, they cannot offer the near-instantaneous feedback that fully-automated assessment approaches can offer.
The impracticality of open-ended assignments often leads to instructors designing their assignments to work with traditional auto-grading approaches, limiting the opportunities for students to take control of their learning.

\subsection{Why Assess Maintainability, Readability and Documentation? (RQ3)}
While most tools focus on assessing correctness, few focus on assessing code quality aspects, such as maintainability, readability or documentation.
The automated assessment of these skills is typically minimal and is often focused on static analysis to determine the quality of the source code.

While evaluating these skills using static analysis can provide a good indicator, most static analysis tools and metrics are designed to evaluate professional code.
Some metrics that evaluate these areas may not be suitable for all types of novice programming assignments, especially short-form assignments typically used with unit testing.
This could be due to these assignments providing the overall code design, limiting the ability to use maintainability metrics such as Depth of Inheritance Tree or Coupling Between Object Classes \cite{Chidamber1994}.

In the case of documentation, the static analysis tools are typically limited to detecting the presence of comments and that they follow the correct format.
However, recent research has started to investigate how to implement metrics commonly used in prose that can be applied to evaluating documentation.
\citet{Eleyan2020} use the Flesch reading ease score and Flesch-Kincaid grade level to evaluate how understandable comments and docstrings are and how long they take to read.

Most AATs that assess these skills using static analysis typically output the result, either as a number or an indication that something is missing.
They do not typically tailor the output of these professional software engineering tools to novice programmers by adapting the result to something a novice programmer can use to improve.
The limited information provided to the student can be confusing, limit progress, and frustrate students \cite{Denny2020}.

While assessing correctness evaluates students' ability to write working code, assessing maintainability, readability, and documentation evaluate their ability to write good code.
Typically, source code that is maintainable, readable and well-documented is easier to adapt, especially when working in teams or for future development.
Furthermore, evaluating students' adherence to consistent code style is critical in teaching students to write readable and well-documented code \cite{Hart2023}.

\subsection{Degree of Automation (RQ1)}
There are advantages and disadvantages to both fully automated and semi-automated assessment.
Fully-automated assessment allows for near-instantaneous feedback but typically limits the scope of the assignment.
Providing near-instantaneous feedback can encourage students to submit early \cite{Leinonen2022} and address underlying misconceptions \cite{Gusukuma2018}. 

Most fully-automated assessment tools use unit testing to assess student solutions, but they share similar issues as most unit-test-based AATs.
These include forcing instructors to design closed-ended or structured assignments compatible with the tool and limiting the opportunities for students to take control of their assignments. %
Additionally, maintainability, readability and documentation assessment is often non-existent or limited static analysis, such as conforming to a style guide or matching specific patterns.
The typical structure used in these AATs makes it highly challenging to automatically grade elements such as correct use of object-oriented concepts, class and variable naming, and documentation quality, as these are typically provided by starter code designed to work with the AAT.

While semi-automated assessment does not allow for near-instantaneous feedback, it allows feedback to be delivered faster than manual assessment.
Semi-automated assessment typically aids the grader by assessing the correctness of the submission and allowing the human grader to assess the other skills.
Providing feedback quickly is essential to high student satisfaction \cite{Kane2008}.
However, the typical use of semi-automated assessment does not resolve the issues around close-ended assignments, especially providing students with a specific structure.
The human grader can manually assess documentation quality, how students have named their local variables and the code style.
But they can still not assess key skills like the correct use of object-oriented principles, as this is typically defined in the provided starter code.

One potential solution to automatically assess open-ended assignments, at least partially, is to automate the grading of maintainability, readability and documentation and manually grade the correctness.
This semi-automated approach would allow automated grading of elements commonly shared between solutions, such as documentation quality, variable naming, adherence to code style, and code design principles. 
While allowing human graders to focus on assessing the implemented functionality and whether the student's solution met the requirements in the open-ended assignment.
This approach would allow instructors to set open-ended assignments while offering students partial near-instantaneous feedback on skills that are rarely graded automatically.
Furthermore, having the human graders only assess the correctness can reduce the overall assessment time and potentially reduce the variety in grades typically produced when multiple human graders grade the more subjective skills, including maintainability, readability and documentation.

Additionally, further research could investigate the effect of providing the results of the automatically assessed elements near-instantaneously to the student and providing the manually assessed element after the deadline.
For example, students could receive continuous feedback on their maintainability, readability and documentation, and their final grade and feedback when the correctness has been manually assessed, providing students with feedback on areas that typically take time to grade and allowing instructors to set open-ended assignments.

\subsection{Language Paradigms Graded (RQ2)}
While surveys into popular programming languages\footnote{GitHub -- The top programming languages 2022 (accessed 30/01/23): \url{https://octoverse.github.com/2022/top-programming-languages}}\textsuperscript{,}\footnote{JetBrains -- The State of Developer Ecosystem 2022 (accessed 30/01/23): \url{https://www.jetbrains.com/lp/devecosystem-2022/}}\textsuperscript{,}\footnote{StackOverflow -- Developer Survey 2022 (accessed 30/01/23): \url{https://survey.stackoverflow.co/2022/\#technology}} have yielded that JavaScript, a web-based language, is the most commonly used programming language, OOP languages are still very prominent.
Besides providing fundamental skills, the popularity of OOP languages could be why they are the most commonly automatically assessed paradigm.
Additionally, most OOP languages have a framework that can be used for web development in conjunction with web-based languages, such as Spring\footnote{Spring -- A Framework for Java Microservices and Web-Apps: \url{https://spring.io/}} for Java and Django\footnote{Django -- A web framework for Python: \url{https://www.djangoproject.com/}} for Python.

These frameworks could provide a potential route to teaching web development.
Most introductory courses teach an OOP-based language, allowing students to learn server-side-based web development without first needing to learn client-side web-based languages.
Furthermore, these OOP-based frameworks often support unit tests, potentially allowing for existing AATs to be used to assess web-development assignments.
However, further research should be conducted into the automated assessment of web-based languages, as these assessments are rarely automated.

\subsection{Evaluation Techniques}
\subsubsection{Approaches}
Conducting student and instructor surveys has many benefits, including providing insight into the users' experience with the tool, how well it worked for them, and how it could be improved.
However, surveys cannot validate the accuracy of AATs, as the surveys only collect the user's opinion of how well the tool performed.
To validate the accuracy of AATs, the results of the proposed tool should be compared to a benchmark, either other tools or, ideally, a human-graded dataset.

While comparing to a benchmark dataset can validate the tool's accuracy, human graders are typically better at understanding the nuances in a student's submission. This allows them to provide more accurate assessments, especially for partial or uncompilable submissions.
However, using multiple human graders to assess large courses is common practice, which can introduce some variability in awarded grades and feedback given.
Evaluating against other tools can demonstrate improved accuracy when running against the same benchmark.

Conducting a mixture of quantitative evaluation in the form of comparing against benchmarks, both against other tools and human graders, and qualitative evaluation, such as student and instructor surveys, could provide the most in-depth analysis of AATs.
This allows researchers to show how their tool improves upon other tools and compares to the gold standard of human graders while also providing valuable insights into the user's experience, both from a student and instructor's point of view.

\subsubsection{Data Availability}
Publicly distributing datasets would allow instructors and researchers to compare similar and new AATs against a shared dataset, allowing them to make informed decisions on which tools to use or adapt for their purposes instead of developing another automated grader from scratch.
Furthermore, releasing datasets alongside the study allows researchers to reproduce and validate experiment results, aiding instructors in choosing which AAT to use.
Among the released datasets, few are utilised in multiple papers.
Those utilised are typically from large-scale courses using smaller-scale programming exercises from online judges, such as Hacker Rank \cite{rayyan-354359338, rayyan-354359341}.
The lack of validated available benchmarks can make it difficult for researchers to validate the results of their tools, especially for AATs that focus on less commonly assessed skills, including maintainability, readability and documentation. 

In addition to increasing reproducibility and providing data for benchmarking, releasing publicly available datasets could also allow researchers to find relevant data for their research, whether automated assessment-related or other code-based research, without requiring the researchers to develop a new dataset.
This could decrease the overall time spent on a project and produce cutting-edge research faster.

\subsection{Limitations of AAT Provided Feedback (RQ6)}
Most AATs provided feedback in one form or another, with dynamic analysis tools typically showing whether the test cases succeeded and if they failed the expected output and actual output or the exception/compiler message if one was thrown.
Providing the test results can aid students' learning, especially when providing near-instantaneous feedback allowing students to get feedback on their progress.
However, the lack of detailed feedback can frustrate students when they cannot figure out why certain test cases are failing, for example, when the expected and actual outputs look identical to the student but have an unnoticed trailing space.
Furthermore, passing compiler and runtime exceptions directly to the students without post-processing is inadequate and presents a barrier to progress and a source of discouragement \cite{Becker2016}.

Feedback utilising static analysis also shares similar limitations. 
The tools that use code repair typically suggest edits that can be made to make the code compilable but with a limited explanation of why the suggested fix makes the code compile.
Tools based on software metrics or linters often provide overwhelming feedback to the student, typically highlighting each occurrence of a readability, maintainability or documentation issue.
Furthermore, as these AATs are typically based on tools designed for professional software engineers, the feedback supplied to the student can be confusing and contain feedback on topics they have yet to learn about.

While there are limitations to feedback provided by AATs, human-provided feedback is also imperfect.
Instructors can provide more nuanced and directed feedback for particular students. 
However, this takes time, and when assessing large cohorts is practically impossible.
AATs typically provide enough feedback to aid most students' learning while allowing instructors more time to aid struggling students.
Further research into the effects of AATs on student learning compared to assessment by human graders could be undertaken in the future.

\subsection{Performance Against Human Assessors (RQ4, RQ5)}
Evaluating AATs against human assessors is a common method of assessing the quality of AATs, second to conducting user studies.
Most evaluate how accurate or well the AAT grades correlate with human graders, with few comparing the automatically generated and human-generated feedback.
Those that evaluate automatically generated feedback against human-generated feedback primarily focus on whether the instructors agree with the generated feedback provided or if the generated feedback is accurate regarding errors detected or categorisation of issues within the code.

Further research is required to evaluate the learning effect of automatically generated grades and feedback compared to human-provided grades and feedback, mainly which elements are assessed and the quality and quantity of the feedback provided to the students. 

\subsection{Results Compared to Related Systematic Literature Reviews}\label{sect:discussion:related_work}

Most of the related work investigates AATs outside of our publication window of 2017 to 2021; here, we compare our results with the previous reviews in this area to discuss any potential long-term trends when combined with our results.
\citet{Souza2016} investigated AATs between 1985 and 2013 and found that most tools were fully automated, with less than a quarter of tools that they reviewed opting to develop a semi-automated tool.
This trend for fully automated tools has not changed in our review, and we also found that most tools were fully automated, with only 14\% of tools using a semi-automated approach.

Similarly, \citet{Keuning2018} focused on conducting a systematic literature review of feedback provided by AATs between 1960 and 2015.
They provide a more in-depth analysis of feedback provided by AATs by utilising \citeauthor{Narciss2008}'s~\citep{Narciss2008} feedback categories and found that most AATs provide feedback on finding mistakes using test-based feedback.
Our results also show this trend of providing feedback on mistakes, primarily using the output of dynamic analysis.
\citeauthor{Keuning2018} also found that some AATs provide feedback on how to proceed. 
Our review also found instances of AATs providing feedback on how to proceed, typically tools that provide feedback based on code repair.

Furthermore, \citeauthor{Keuning2018} categorised the tools by language paradigm assessed and had similar results to our review, with most tools assessing object-oriented languages and few tools supporting the assessment of logic, functional or other paradigms.
They also analysed the quality of the AATs they reviewed. 
They found that most utilised empirical approaches, such as comparing to learning objectives, conducting student and teacher surveys, or evaluating the AAT based on the time taken to complete a task, with other tools being evaluated analytically, anecdotally or not at all.

We found similar trends in assessing object-oriented languages and most tools being evaluated using empirical methods.
However, we provided a more in-depth analysis of evaluation techniques and found that most tools are evaluated by surveys or by comparing the results against manual grading.

While the other systematic reviews discussed investigate papers published in years that do not overlap our review, \citet{Paiva2022}'s review investigates publications between 2010 and 2021.
They investigated assessment techniques and found that whitebox dynamic analysis and static analysis are gaining more traction as methods to assess the functionality of submissions. 

For code quality and software metrics, they discuss tools that utilise existing code quality tools and software metrics.
While there is an overlap to \citeauthor{Paiva2022}'s review, our review provides a more detailed analysis of the techniques used to assess submissions by introducing categorising tools by both the key skills and the approaches used -- whereas \citeauthor{Paiva2022} primarily focus on the domains graded, such as visual programming, computer graphics, and software testing, how secure the code execution is, and the effectiveness of the tools.

\subsection{Threats to Validity}
Limiting the search for primary studies to IEEExplore, ACM Digital Library, and Scopus may have led to some relevant primary studies not published in an IEEE, ACM or a Scopus-indexed publication being missed.
Additionally, given that some titles and abstracts did not clearly state what their papers addressed, some papers may have been mistakenly excluded during the title and abstract screening.
To mitigate incorrectly excluding papers at this stage, the screeners opted to include such papers in the title and abstract screening to allow further analysis during the introduction and conclusion screening stage.
If, during the screening stage, the screeners disagreed on whether a paper should be included, they discussed their disagreements and came to a consensus if the paper should be included.

We opted not to conduct a snowball search, where after the screening stage, included papers references are searched for any other papers that match the inclusion criteria.
We decided against conducting a snowball search, as we felt that our final number of included papers was enough to provide a practical overview of the state of the art without delaying the publication of our review.
However, as a result, there are other papers we most likely have missed in our initial search that do not contain the keywords in our search string in their titles or abstracts.

Commercial AATs may not have been included in our study as they do not have any associated academic literature.
While many institutions use these commercial tools, the lack of literature discussing them makes them out of scope for our literature review.

Although both screeners conducted the first round of annotations during the full-text screening stage, additional annotations were needed while extracting the results.
These additional annotations were compiled by a single screener, with the final list of applied annotations being validated by the second screener.
Therefore, some papers may have been mislabeled or missing relevant annotations.

While writing our results for this paper, there was a rise in research into using large language models (LLMs) within computer science education.
This rise in LLMs can affect automated assessment tools, especially with smaller, constrained assignments that are more susceptible to the code generation functionality of LLMs.
One study found that if ChatGPT is given clear and straightforward instructions, it can generate effective solutions for trivial and constrained assignments \cite{Ouh2023}.
Another study found that the solutions obtained for non-trivial programming assignments are not sufficient for the competition of the course, the LLM can correct solutions based on feedback from an AAT \cite{Savelka2023}.

\section{Conclusion}\label{sect:conculusion}
This systematic review categorised state-of-the-art automated grading and feedback tools by the graded programming skills, techniques for awarding a grade or generating feedback, programming paradigms for automatically assessing programming assignments, and how these tools were evaluated.
Furthermore, we investigated why automated assessment tools focus on assessing specific programming fundamental skills, discussed potential reasons for choosing a particular language and degree of automation, and investigated how researchers evaluated tools and how evaluation could be improved.

We found that most AATs in the scope of this review focus on assessing correctness using a dynamic approach, most of which used unit testing to provide the grade and feedback.
Feedback is typically limited to whether the unit test has passed or failed and the expected and actual outcomes if the test has failed.
This can leave students frustrated, as often the feedback does not provide enough detail to help the student to progress.

Another common approach to assess correctness is a static approach comparing a student's submission to a reference solution or a set of correct student submissions.
Static analysis is also used to assess both maintainability, readability and the presence of documentation.
However, these skills are assessed less often and typically in conjunction with correctness grading.

Instructors focus on assessing correctness, which is typically seen as one of the most crucial skills. 
They can determine if students have used a specific language feature or algorithm and if students can understand and convert requirements into a complete code base.
While correctness is often seen as one of the most crucial skills, maintainability, readability and documentation are also vital.
Maintainable, readable and well-documented code is typically easier to develop further, especially when working in teams or on future releases.

Most tools offered fully automated assessment, allowing for near-instantaneous feedback and multiple resubmissions without increasing the grading workload.
Receiving feedback quickly increases student satisfaction, and multiple submissions allow students more opportunities to succeed.
However, fully automated tools typically limit the scope of the assignment to a smaller scale and limit opportunities to show creativity.

Some tools opt for semi-automated approaches, where human graders manually assess elements of the assignment, typically maintainability, readability and documentation, and automatically assess correctness.
While semi-automated approaches do not allow for near-instantaneous feedback, they are faster than manual assessment.
However, typical implementations do not resolve the issues around limiting the scope of assignments, as correctness is typically assessed using the same methodology as full-automated assessment.

In terms of language paradigms assessed, most assess object-oriented languages, such as Java, Python and C++. 
Other language paradigms, such as functional and logic languages, have AATs, but these are researched less often.
Object-oriented languages are the primary focus for many automated assessments due to the prominence of these languages in education and industry.
Other language paradigms are gaining more popularity, especially web-based languages like JavaScript.

Most papers evaluate an AAT, whether one they have developed or used in a course.
The primary evaluation technique was to conduct student surveys about their thoughts on the tool and how using the tool aided their learning.
Another common evaluation technique was to compare the AAT to human graders, most focusing on the accuracy of the assessment using the human graders' marks as a benchmark.
The dataset used for evaluation is typically not published.
While most papers perform some form of evaluation, the evaluation is typically focused on a single tool, is conducted by the tool's developer and has mostly positive results.

While evaluating tools with student and teacher surveys can provide valuable insights into the users' experience with the tools, they do not evaluate the accuracy of the awarded grades and feedback.
Evaluating the accuracy against a benchmark or to human graders in conjunction with the user's experience allows instructors to compare similar tools for considerations in their courses.
Releasing evaluation datasets would allow researchers to reproduce and validate experiment results and evaluate different tools against a common set of assignments. 
This would improve the evaluation by providing verifiable and comparable benchmarks for the tool's accuracy. 

\subsection{Recommendations}
With most research into automated assessment of programming assignments focusing on assessing correctness for small-scale closed assignments, we encourage researchers to investigate how to automatically assess maintainability, readability and documentation, as these are key skills that are not evaluated by most automated assessment tools.
Furthermore, we suggest that future research investigates how to assess open-ended assignments.
Including semi-automated approaches to automating the assessment of maintainability, readability and documentation while manually assessing correctness or designing open-ended assignments to automate the assessment fully.

As web-based languages become more prominent, future research could investigate the automatic assessment of web-based languages, including JavaScript and TypeScript.
In addition to web-based languages, automatic assessment of web application development could be investigated further by investigating how to assess the use of popular web frameworks, user-experience design and web-testing frameworks.

While researching new methods of automatic assessment for open-ended assignments, maintainability, readability and documentation and web-based languages, we suggest that authors produce more robust evaluation practices and attempt to publish the datasets they used for their evaluation.
While some tools are evaluated by comparing their results to human graders, most are only evaluated by student or teacher surveys or performance analytics, such as compute power required or run times.
Evaluating tools with a survey can provide meaningful insights into the end-users opinions of a tool, but they cannot adequately determine the accuracy of the tool.
The publication of annotated datasets would allow researchers to evaluate a tool's accuracy to similar tools on the same dataset, providing more significant evidence of the tool's performance.
Evaluating against a benchmark dataset and using user surveys could provide a good mix of qualitative and quantitative evidence to support the performance of their tools.

\subsection{Data Availability}
Our final list of annotated papers and data processing pipeline is available on GitHub\footnote{Raw data and data processing repository: \url{https://github.com/m-messer/Automated-Assessment-SLR-Data-Processing}}.

\begin{acks}
We appreciate the reviewers' insightful remarks and recommendations.
\end{acks}

\printbibliography[notkeyword={results}]
\printbibliography[keyword={results},title={Publications Included as Part of the Search Process}]

\appendix
\section{Supplementary Material}
\subsection{Publications by Skill Graded and Category of Automatic Assessment Tools}

\begin{table}[H]
\footnotesize
\begin{tabular}{|l|l|p{0.23\textwidth}|p{0.23\textwidth}|p{0.23\textwidth}|}
\hline
\multicolumn{2}{|l|}{} & \multicolumn{3}{c|}{\textbf{Category}} \\ \cline{3-5} 
\multicolumn{2}{|l|}{} & Dynamic & Static & Machine Learning \\ \hline
\multirow{20}{*}{\rotatebox[origin=c]{90}{\textbf{Skill}}} & Correctness & \citep{rayyan-354359269, rayyan-354359364, rayyan-354359317, rayyan-354359316, rayyan-354359315, rayyan-354359314, rayyan-354359313, rayyan-354359311, rayyan-354359366, rayyan-354359309, rayyan-354359368, rayyan-354359308, rayyan-354359307, rayyan-354359370, rayyan-354359372, rayyan-354359373, rayyan-354359376, rayyan-354359377, rayyan-354359318, rayyan-354359338, rayyan-354359319, rayyan-354359321, rayyan-354359336, rayyan-354359339, rayyan-354359334, rayyan-354359340, rayyan-354359333, rayyan-354359342, rayyan-354359345, rayyan-354359331, rayyan-354359348, rayyan-354359330, rayyan-354359350, rayyan-354359353, rayyan-354359354, rayyan-354359326, rayyan-354359355, rayyan-354359359, rayyan-354359363, rayyan-354359320, rayyan-354359301, rayyan-354359302, rayyan-354359379, rayyan-354359401, rayyan-354359280, rayyan-354359403, rayyan-354359279, rayyan-354359404, rayyan-354359278, rayyan-354359277, rayyan-354359406, rayyan-354359276, rayyan-354359407, rayyan-354359408, rayyan-354359273, rayyan-354359272, rayyan-354359410, rayyan-354359411, rayyan-354359412, rayyan-354359270, rayyan-354359282, rayyan-354359300, rayyan-354359284, rayyan-354359285, rayyan-354359381, rayyan-354359385, rayyan-354359386, rayyan-354359387, rayyan-354359296, rayyan-354359390, rayyan-354359391, rayyan-354359397, rayyan-354359294, rayyan-354359392, rayyan-354359291, rayyan-354359394, rayyan-354359290, rayyan-354359395, rayyan-354359287, rayyan-354359396, rayyan-354359293, rayyan-354359337, rayyan-354359367} & \citep{rayyan-354359388, rayyan-354359297, rayyan-354359298, rayyan-354359291, rayyan-354359336, rayyan-354359302, rayyan-354359376, rayyan-354359303, rayyan-354359375, rayyan-354359304, rayyan-354359305, rayyan-354359299, rayyan-354359383, rayyan-354359392, rayyan-354359280, rayyan-354359279, rayyan-354359278, rayyan-354359405, rayyan-354359274, rayyan-354359407, rayyan-354359273, rayyan-354359408, rayyan-354359272, rayyan-354359409, rayyan-354359410, rayyan-354359411, rayyan-354359270, rayyan-354359290, rayyan-354359367, rayyan-354359414, rayyan-354359366, rayyan-354359357, rayyan-354359327, rayyan-354359323, rayyan-354359309, rayyan-354359349, rayyan-354359347, rayyan-354359361, rayyan-354359346, rayyan-354359363, rayyan-354359317, rayyan-354359344, rayyan-354359332, rayyan-354359341, rayyan-354359334, rayyan-354359322, rayyan-354359325} & \citep{rayyan-354359367, rayyan-354359369, rayyan-354359409, rayyan-354359341, rayyan-354359310, rayyan-354359285, rayyan-354359328, rayyan-354359295} \\ \cline{2-5} 
  & Maintainability & \citep{rayyan-354359354, rayyan-354359377} & \citep{rayyan-354359403, rayyan-354359413, rayyan-354359412, rayyan-354359271, rayyan-354359276, rayyan-354359405, rayyan-354359351, rayyan-354359378, rayyan-354359352, rayyan-354359383, rayyan-354359298} &  \\ \cline{2-5} 
  & Readability &  & \citep{rayyan-354359386, rayyan-354359354, rayyan-354359308, rayyan-354359333, rayyan-354359387, rayyan-354359413, rayyan-354359271, rayyan-354359368, rayyan-354359306, rayyan-354359344, rayyan-354359331, rayyan-354359362, rayyan-354359391, rayyan-354359360, rayyan-354359404, rayyan-354359330, rayyan-354359383, rayyan-354359337, rayyan-354359328, rayyan-354359398, rayyan-354359284, rayyan-354359396, rayyan-354359294, rayyan-354359287, rayyan-354359394, rayyan-354359351, rayyan-354359352, rayyan-354359297} & \cite{rayyan-354359369} \\ \cline{2-5} 
  & Documentation &  & \cite{rayyan-354359305} &  \\ \hline
\end{tabular}
\end{table}

\subsection{Publications by Technique Implemented}
\footnotesize

\begin{table}[H]
\begin{tabular}{|ll|l|p{0.65\textwidth}|}
\hline
\multicolumn{2}{|l|}{} & \textbf{Technique} & \textbf{References} \\ \hline
\multicolumn{1}{|l|}{\multirow{22}{*}{\rotatebox[origin=c]{90}{\textbf{Category}}}} & \multirow{11}{*}{\rotatebox[origin=c]{90}{Static}} & Static Analysis & \cite{rayyan-354359271, rayyan-354359278, rayyan-354359279, rayyan-354359284, rayyan-354359294, rayyan-354359305, rayyan-354359306, rayyan-354359328, rayyan-354359330, rayyan-354359331, rayyan-354359333, rayyan-354359337, rayyan-354359339, rayyan-354359344, rayyan-354359345, rayyan-354359360, rayyan-354359366, rayyan-354359368, rayyan-354359378, rayyan-354359383, rayyan-354359386, rayyan-354359394, rayyan-354359398, rayyan-354359403, rayyan-354359404, rayyan-354359405, rayyan-354359408, rayyan-354359410, rayyan-354359412, rayyan-354359413} \\ \cline{3-4} 
\multicolumn{1}{|l|}{} &  & Pattern Matching & \citep{rayyan-354359270, rayyan-354359272, rayyan-354359273, rayyan-354359274, rayyan-354359278, rayyan-354359279, rayyan-354359298, rayyan-354359303, rayyan-354359309, rayyan-354359322, rayyan-354359323, rayyan-354359325, rayyan-354359327, rayyan-354359328, rayyan-354359332, rayyan-354359341, rayyan-354359347, rayyan-354359349, rayyan-354359351, rayyan-354359361, rayyan-354359375, rayyan-354359409, rayyan-354359414} \\ \cline{3-4} 
\multicolumn{1}{|l|}{} &  & Model Solution Required & \citep{rayyan-354359270, rayyan-354359274, rayyan-354359276, rayyan-354359278, rayyan-354359291, rayyan-354359297, rayyan-354359298, rayyan-354359320, rayyan-354359323, rayyan-354359328, rayyan-354359332, rayyan-354359336, rayyan-354359341, rayyan-354359347, rayyan-354359349, rayyan-354359350, rayyan-354359361, rayyan-354359375, rayyan-354359378, rayyan-354359397, rayyan-354359401, rayyan-354359412} \\ \cline{3-4} 
\multicolumn{1}{|l|}{} &  & Style Check & \citep{rayyan-354359287, rayyan-354359294, rayyan-354359305, rayyan-354359308, rayyan-354359318, rayyan-354359336, rayyan-354359354, rayyan-354359360, rayyan-354359362, rayyan-354359363, rayyan-354359366, rayyan-354359368, rayyan-354359376, rayyan-354359391, rayyan-354359396, rayyan-354359403, rayyan-354359404} \\ \cline{3-4} 
\multicolumn{1}{|l|}{} &  & Program Repair & \citep{rayyan-354359271, rayyan-354359279, rayyan-354359290, rayyan-354359299, rayyan-354359304, rayyan-354359330, rayyan-354359334, rayyan-354359346, rayyan-354359357, rayyan-354359361, rayyan-354359367, rayyan-354359375} \\ \cline{3-4} 
\multicolumn{1}{|l|}{} &  & Model Solution Closeness & \citep{rayyan-354359274, rayyan-354359278, rayyan-354359279, rayyan-354359298, rayyan-354359299, rayyan-354359325, rayyan-354359328, rayyan-354359332, rayyan-354359341, rayyan-354359346, rayyan-354359347} \\ \cline{3-4} 
\multicolumn{1}{|l|}{} &  & Code Repair for Feedback & \citep{rayyan-354359270, rayyan-354359299, rayyan-354359303, rayyan-354359304, rayyan-354359346, rayyan-354359367, rayyan-354359375, rayyan-354359401} \\ \cline{3-4} 
\multicolumn{1}{|l|}{} &  & Source Code Metrics & \citep{rayyan-354359271, rayyan-354359297, rayyan-354359298, rayyan-354359344, rayyan-354359383, rayyan-354359403, rayyan-354359413} \\ \cline{2-4} 
\multicolumn{1}{|l|}{} & \multirow{7}{*}{\rotatebox[origin=c]{90}{Dynamic}} & Unit Testing &  \citep{rayyan-354359269, rayyan-354359270, rayyan-354359272, rayyan-354359273, rayyan-354359276, rayyan-354359277, rayyan-354359278, rayyan-354359279, rayyan-354359280, rayyan-354359282, rayyan-354359284, rayyan-354359285, rayyan-354359287, rayyan-354359290, rayyan-354359291, rayyan-354359293, rayyan-354359294, rayyan-354359296, rayyan-354359300, rayyan-354359301, rayyan-354359302, rayyan-354359307, rayyan-354359308, rayyan-354359309, rayyan-354359311, rayyan-354359313, rayyan-354359314, rayyan-354359315, rayyan-354359316, rayyan-354359317, rayyan-354359318, rayyan-354359319, rayyan-354359320, rayyan-354359321, rayyan-354359326, rayyan-354359330, rayyan-354359331, rayyan-354359333, rayyan-354359334, rayyan-354359336, rayyan-354359337, rayyan-354359338, rayyan-354359339, rayyan-354359340, rayyan-354359342, rayyan-354359345, rayyan-354359348, rayyan-354359350, rayyan-354359353, rayyan-354359354, rayyan-354359355, rayyan-354359359, rayyan-354359363, rayyan-354359364, rayyan-354359366, rayyan-354359368, rayyan-354359370, rayyan-354359372, rayyan-354359373, rayyan-354359376, rayyan-354359377, rayyan-354359379, rayyan-354359381, rayyan-354359385, rayyan-354359386, rayyan-354359387, rayyan-354359390, rayyan-354359391, rayyan-354359392, rayyan-354359394, rayyan-354359395, rayyan-354359396, rayyan-354359397, rayyan-354359401, rayyan-354359403, rayyan-354359404, rayyan-354359406, rayyan-354359407, rayyan-354359408, rayyan-354359410, rayyan-354359411}\\ \cline{3-4} 
\multicolumn{1}{|l|}{} &  & Output Matching & \cite{rayyan-354359272, rayyan-354359279, rayyan-354359310, rayyan-354359322, rayyan-354359326, rayyan-354359327, rayyan-354359372} \\ \cline{3-4} 
\multicolumn{1}{|l|}{} &  & Property-Based Testing & \citep{rayyan-354359314, rayyan-354359355, rayyan-354359412} \\ \cline{2-4} 
\multicolumn{1}{|l|}{} & \multirow{4}{*}{\rotatebox[origin=c]{90}{Other}} & Machine Learning & \citep{rayyan-354359285, rayyan-354359295, rayyan-354359310, rayyan-354359328, rayyan-354359330, rayyan-354359341, rayyan-354359367, rayyan-354359369, rayyan-354359409, rayyan-354359413} \\ \cline{3-4} 
\multicolumn{1}{|l|}{} &  & Other & \citep{rayyan-354359351, rayyan-354359293, rayyan-354359313, rayyan-354359348, rayyan-354359362, rayyan-354359363, rayyan-354359385, rayyan-354359408, rayyan-354359299, rayyan-354359303, rayyan-354359346, rayyan-354359362, rayyan-354359274, rayyan-354359299, rayyan-354359278, rayyan-354359279, rayyan-354359294, rayyan-354359361, rayyan-354359285, rayyan-354359290, rayyan-354359325, rayyan-354359332, rayyan-354359310, rayyan-354359340, rayyan-354359325, rayyan-354359407, rayyan-354359305, rayyan-354359336, rayyan-354359337, rayyan-354359276, rayyan-354359280, rayyan-354359282, rayyan-354359291, rayyan-354359297, rayyan-354359298, rayyan-354359302, rayyan-354359305, rayyan-354359308, rayyan-354359317, rayyan-354359323, rayyan-354359325, rayyan-354359351, rayyan-354359352, rayyan-354359354, rayyan-354359387, rayyan-354359388, rayyan-354359392, rayyan-354359405, rayyan-354359407, rayyan-354359411, rayyan-354359412, rayyan-354359276, rayyan-354359316, rayyan-354359354, rayyan-354359377, rayyan-354359328, rayyan-354359320, rayyan-354359269, rayyan-354359352, rayyan-354359354, rayyan-354359387, rayyan-354359413, rayyan-354359397, rayyan-354359290, rayyan-354359297, rayyan-354359320, rayyan-354359366, rayyan-354359322, rayyan-354359277, rayyan-354359313, rayyan-354359347, rayyan-354359349, rayyan-354359361, rayyan-354359394, rayyan-354359307} \\ \hline
\end{tabular}
\end{table}

\subsection{Publications by Degree of Automation}
\footnotesize
\begin{table}[H]
\begin{tabular}{|l|p{0.75\textwidth}|}
\hline
\textbf{Degree of Automation} & \textbf{References} \\ \hline
Fully Automated & \citep{rayyan-354359269, rayyan-354359369, rayyan-354359368, rayyan-354359367, rayyan-354359366, rayyan-354359364, rayyan-354359363, rayyan-354359362, rayyan-354359361, rayyan-354359360, rayyan-354359359, rayyan-354359357, rayyan-354359370, rayyan-354359355, rayyan-354359353, rayyan-354359352, rayyan-354359351, rayyan-354359350, rayyan-354359349, rayyan-354359347, rayyan-354359346, rayyan-354359345, rayyan-354359344, rayyan-354359341, rayyan-354359413, rayyan-354359354, rayyan-354359372, rayyan-354359373, rayyan-354359375, rayyan-354359412, rayyan-354359411, rayyan-354359410, rayyan-354359409, rayyan-354359408, rayyan-354359404, rayyan-354359403, rayyan-354359401, rayyan-354359398, rayyan-354359397, rayyan-354359396, rayyan-354359395, rayyan-354359394, rayyan-354359392, rayyan-354359391, rayyan-354359390, rayyan-354359388, rayyan-354359387, rayyan-354359386, rayyan-354359385, rayyan-354359381, rayyan-354359379, rayyan-354359378, rayyan-354359377, rayyan-354359376, rayyan-354359339, rayyan-354359338, rayyan-354359340, rayyan-354359303, rayyan-354359307, rayyan-354359305, rayyan-354359336, rayyan-354359302, rayyan-354359301, rayyan-354359300, rayyan-354359299, rayyan-354359298, rayyan-354359297, rayyan-354359296, rayyan-354359308, rayyan-354359294, rayyan-354359285, rayyan-354359284, rayyan-354359282, rayyan-354359279, rayyan-354359278, rayyan-354359277, rayyan-354359276, rayyan-354359274, rayyan-354359273, rayyan-354359271, rayyan-354359293, rayyan-354359309, rayyan-354359414, rayyan-354359311, rayyan-354359330, rayyan-354359310, rayyan-354359325, rayyan-354359323, rayyan-354359331, rayyan-354359322, rayyan-354359326, rayyan-354359320, rayyan-354359327, rayyan-354359317, rayyan-354359332, rayyan-354359315, rayyan-354359333, rayyan-354359334, rayyan-354359319} \\ \hline
Semi-Automated & \citep{rayyan-354359407, rayyan-354359405, rayyan-354359272, rayyan-354359342, rayyan-354359270, rayyan-354359280, rayyan-354359406, rayyan-354359337, rayyan-354359287, rayyan-354359291, rayyan-354359295, rayyan-354359383, rayyan-354359321, rayyan-354359318, rayyan-354359314, rayyan-354359313, rayyan-354359348} \\ \hline
Unclear & \citep{rayyan-354359328, rayyan-354359316, rayyan-354359304, rayyan-354359306, rayyan-354359290} \\ \hline
\end{tabular}
\end{table}

\subsection{Publications by Language Paradigm}
\footnotesize
\begin{table}[H]
\begin{tabular}{|l|p{0.75\textwidth}|}
\hline
\textbf{Language Paradigm} & \textbf{References} \\ \hline
OOP & \citep{rayyan-354359355, rayyan-354359396, rayyan-354359353, rayyan-354359352, rayyan-354359351, rayyan-354359350, rayyan-354359398, rayyan-354359404, rayyan-354359405, rayyan-354359348, rayyan-354359347, rayyan-354359346, rayyan-354359408, rayyan-354359345, rayyan-354359344, rayyan-354359410, rayyan-354359411, rayyan-354359357, rayyan-354359359, rayyan-354359387, rayyan-354359361, rayyan-354359385, rayyan-354359383, rayyan-354359381, rayyan-354359379, rayyan-354359378, rayyan-354359377, rayyan-354359376, rayyan-354359391, rayyan-354359340, rayyan-354359392, rayyan-354359394, rayyan-354359370, rayyan-354359369, rayyan-354359367, rayyan-354359364, rayyan-354359363, rayyan-354359395, rayyan-354359388, rayyan-354359360, rayyan-354359269, rayyan-354359413, rayyan-354359306, rayyan-354359339, rayyan-354359303, rayyan-354359302, rayyan-354359300, rayyan-354359299, rayyan-354359298, rayyan-354359297, rayyan-354359296, rayyan-354359295, rayyan-354359294, rayyan-354359307, rayyan-354359293, rayyan-354359290, rayyan-354359287, rayyan-354359285, rayyan-354359284, rayyan-354359282, rayyan-354359279, rayyan-354359277, rayyan-354359276, rayyan-354359273, rayyan-354359271, rayyan-354359270, rayyan-354359291, rayyan-354359308, rayyan-354359305, rayyan-354359322, rayyan-354359330, rayyan-354359328, rayyan-354359326, rayyan-354359325, rayyan-354359323, rayyan-354359309, rayyan-354359321, rayyan-354359320, rayyan-354359319, rayyan-354359333, rayyan-354359318, rayyan-354359316, rayyan-354359315, rayyan-354359314, rayyan-354359313, rayyan-354359337, rayyan-354359311, rayyan-354359310, rayyan-354359317, rayyan-354359334} \\ \hline
Functional & \citep{rayyan-354359336, rayyan-354359327, rayyan-354359403, rayyan-354359375, rayyan-354359401, rayyan-354359397, rayyan-354359354, rayyan-354359314, rayyan-354359307, rayyan-354359412} \\ \hline
Low-Level & \citep{rayyan-354359373, rayyan-354359366, rayyan-354359390} \\ \hline
Web & \citep{rayyan-354359386, rayyan-354359298, rayyan-354359362} \\ \hline
Query & \citep{rayyan-354359278} \\ \hline
Graphics & \citep{rayyan-354359340, rayyan-354359349, rayyan-354359291, rayyan-354359280}\\ \hline
Logic & \citep{rayyan-354359414} \\ \hline
Agnostic & \citep{rayyan-354359339, rayyan-354359372, rayyan-354359326, rayyan-354359406, rayyan-354359332, rayyan-354359301, rayyan-354359409} \\ \hline
Unknown & \citep{rayyan-354359342, rayyan-354359272, rayyan-354359341, rayyan-354359274, rayyan-354359368, rayyan-354359331, rayyan-354359304, rayyan-354359407, rayyan-354359338} \\ \hline
\end{tabular}
\end{table}

\subsection{Publications by Evaluation Technique}
\begin{table}[H]
\footnotesize
\begin{tabular}{|l|p{0.75\textwidth}|}
\hline
\textbf{Evaluation Technique} & \textbf{References} \\ \hline
Student Survey & \citep{rayyan-354359293, rayyan-354359350, rayyan-354359385, rayyan-354359386, rayyan-354359319, rayyan-354359387, rayyan-354359388, rayyan-354359373, rayyan-354359391, rayyan-354359392, rayyan-354359349, rayyan-354359277, rayyan-354359395, rayyan-354359313, rayyan-354359396, rayyan-354359314, rayyan-354359381, rayyan-354359284, rayyan-354359378, rayyan-354359397, rayyan-354359412, rayyan-354359355, rayyan-354359301, rayyan-354359413, rayyan-354359340, rayyan-354359366, rayyan-354359311, rayyan-354359273, rayyan-354359368, rayyan-354359370, rayyan-354359297, rayyan-354359352, rayyan-354359375, rayyan-354359406, rayyan-354359376, rayyan-354359296, rayyan-354359369, rayyan-354359346} \\ \hline
Manual Grading & \citep{rayyan-354359334, rayyan-354359357, rayyan-354359407, rayyan-354359363, rayyan-354359406, rayyan-354359405, rayyan-354359383, rayyan-354359340, rayyan-354359364, rayyan-354359366, rayyan-354359345, rayyan-354359360, rayyan-354359367, rayyan-354359397, rayyan-354359341, rayyan-354359333, rayyan-354359269, rayyan-354359414, rayyan-354359302, rayyan-354359300, rayyan-354359299, rayyan-354359332, rayyan-354359297, rayyan-354359294, rayyan-354359282, rayyan-354359303, rayyan-354359279, rayyan-354359278, rayyan-354359310, rayyan-354359321, rayyan-354359272, rayyan-354359322, rayyan-354359323, rayyan-354359326, rayyan-354359270, rayyan-354359274} \\ \hline
Analytics &  \citep{rayyan-354359299, rayyan-354359315, rayyan-354359291, rayyan-354359307, rayyan-354359387, rayyan-354359348, rayyan-354359284, rayyan-354359396, rayyan-354359331, rayyan-354359377, rayyan-354359344, rayyan-354359398, rayyan-354359360, rayyan-354359306, rayyan-354359325, rayyan-354359339, rayyan-354359410, rayyan-354359305, rayyan-354359362, rayyan-354359311}\\ \hline
Auto Grading & \citep{rayyan-354359314, rayyan-354359297, rayyan-354359311, rayyan-354359347, rayyan-354359309, rayyan-354359276, rayyan-354359293, rayyan-354359270, rayyan-354359409, rayyan-354359385, rayyan-354359390, rayyan-354359272, rayyan-354359315, rayyan-354359328, rayyan-354359273} \\ \hline
None & \citep{rayyan-354359394, rayyan-354359285, rayyan-354359287, rayyan-354359271, rayyan-354359298, rayyan-354359411, rayyan-354359280, rayyan-354359330, rayyan-354359351, rayyan-354359342, rayyan-354359354} \\ \hline
User Study & \citep{rayyan-354359401, rayyan-354359395, rayyan-354359273, rayyan-354359364, rayyan-354359361, rayyan-354359310} \\ \hline
Grading Accuracy & \citep{rayyan-354359320, rayyan-354359279, rayyan-354359316, rayyan-354359278, rayyan-354359295} \\ \hline
Teacher Survey & \citep{rayyan-354359337, rayyan-354359304, rayyan-354359296, rayyan-354359376, rayyan-354359352, rayyan-354359350} \\ \hline
Compared to Other Tools & \citep{rayyan-354359361, rayyan-354359290, rayyan-354359372, rayyan-354359375} \\ \hline
Anecdotal & \citep{rayyan-354359359, rayyan-354359403, rayyan-354359317, rayyan-354359353, rayyan-354359318} \\ \hline
Other & \citep{rayyan-354359398, rayyan-354359381, rayyan-354359336, rayyan-354359338, rayyan-354359408, rayyan-354359306, rayyan-354359299, rayyan-354359277, rayyan-354359379, rayyan-354359328, rayyan-354359404, rayyan-354359308, rayyan-354359355, rayyan-354359318} \\ \hline
\end{tabular}
\end{table}

\subsection{Publications by Dataset Availability}
\footnotesize
\begin{table}[H]
\begin{tabular}{|l|p{0.75\textwidth}|}
\hline
\textbf{Dataset Availability} & \textbf{References} \\ \hline
Unavailable & \citep{rayyan-354359269, rayyan-354359368, rayyan-354359366, rayyan-354359364, rayyan-354359363, rayyan-354359362, rayyan-354359361, rayyan-354359360, rayyan-354359359, rayyan-354359355, rayyan-354359354, rayyan-354359353, rayyan-354359352, rayyan-354359351, rayyan-354359350, rayyan-354359349, rayyan-354359348, rayyan-354359346, rayyan-354359344, rayyan-354359342, rayyan-354359413, rayyan-354359339, rayyan-354359337, rayyan-354359336, rayyan-354359369, rayyan-354359370, rayyan-354359372, rayyan-354359373, rayyan-354359412, rayyan-354359411, rayyan-354359409, rayyan-354359408, rayyan-354359407, rayyan-354359406, rayyan-354359405, rayyan-354359404, rayyan-354359403, rayyan-354359398, rayyan-354359397, rayyan-354359334, rayyan-354359395, rayyan-354359392, rayyan-354359391, rayyan-354359390, rayyan-354359388, rayyan-354359387, rayyan-354359386, rayyan-354359385, rayyan-354359381, rayyan-354359379, rayyan-354359377, rayyan-354359376, rayyan-354359394, rayyan-354359333, rayyan-354359340, rayyan-354359331, rayyan-354359304, rayyan-354359303, rayyan-354359332, rayyan-354359301, rayyan-354359300, rayyan-354359298, rayyan-354359297, rayyan-354359296, rayyan-354359295, rayyan-354359294, rayyan-354359293, rayyan-354359291, rayyan-354359290, rayyan-354359287, rayyan-354359285, rayyan-354359284, rayyan-354359282, rayyan-354359280, rayyan-354359279, rayyan-354359274, rayyan-354359273, rayyan-354359272, rayyan-354359271, rayyan-354359305, rayyan-354359306, rayyan-354359302, rayyan-354359308, rayyan-354359330, rayyan-354359327, rayyan-354359326, rayyan-354359323, rayyan-354359322, rayyan-354359307, rayyan-354359321, rayyan-354359320, rayyan-354359319, rayyan-354359318, rayyan-354359317, rayyan-354359414, rayyan-354359311, rayyan-354359310, rayyan-354359316, rayyan-354359313, rayyan-354359309, rayyan-354359315, rayyan-354359314} \\ \hline
Available & \citep{rayyan-354359375, rayyan-354359357, rayyan-354359367, rayyan-354359338, rayyan-354359396, rayyan-354359378, rayyan-354359401, rayyan-354359278, rayyan-354359328, rayyan-354359341} \\ \hline
On Request & \citep{rayyan-354359270, rayyan-354359410, rayyan-354359383, rayyan-354359277, rayyan-354359325, rayyan-354359345, rayyan-354359299, rayyan-354359276, rayyan-354359347} \\ \hline
\end{tabular}
\end{table}

\end{document}